\documentclass{article}  
\usepackage{natbib}    
\usepackage{epsf1990}  

\bibpunct[,]{[}{]}{;}{a}{}{,}

\input psfig.sty
\def\affilsize{\normalsize}

\newcommand\subtitlestring{\ }
\newcommand\subauthorstring{\ }
\newcommand\subaddressstring{\ }
\newcommand\subtitle[1]{\renewcommand\subtitlestring{#1}}
\newcommand\subauthor[1]{\renewcommand\subauthorstring{#1}}
\newcommand\subaddress[1]{\renewcommand\subaddressstring{#1}}

\newcommand\ctpheader{{\small {\em Cosmological Topology in Paris 1998,} 
14 December 1998, Obs. de Paris, eds 
{\em V.~Blanl{\oe}il \& B.F.~Roukema}}}

\newcommand\submaketitle{ 
\twocolumn[ 
\ctpheader
 \\{\ }
\begin{center}
{
{\LARGE 
\section[\subtitlestring, \ \  {\em \subauthorstring}]{\subtitlestring}{\ }
} 
{\large \subauthorstring}\\{\ }\\
{\large \subaddressstring}\\{\ }\\
}
\end{center}]
}

\def\frtoday{Le\space\number\day\space\ifcase\month\or
  janvier\or f\'evrier\or mars\or avril\or mai\or juin\or
  juillet\or ao\^ut\or septembre\or octobre\or 
  novembre\or d\'ecembre\fi\space \number\year}

\usepackage{epsfig}

\title{\ctpheader\\{\ }\\
{\em Cosmological Topology in Paris 1998}
\\Topologie cosmologique \`a Paris 1998
}

\author{Editors: 
Vincent Blanl{\oe}il$^1$
 \& 
Boudewijn F.~Roukema$^{2}$ 
\\
 \affilsize $^1$Institut de recherche math\'ematique avanc\'ee, 
 \affilsize Universit\'e Louis Pasteur et CNRS,\\ 
 \affilsize 7 rue Ren\'e-Descartes, F-67084 Strasbourg Cedex, France\\
 \affilsize
 $^2$Inter-University Centre for Astronomy and Astrophysics, \\
 \affilsize
 Post Bag 4, Ganeshkhind, Pune, 411 007, India (boud@iucaa.ernet.in)\\
 \affilsize {\em blanloei@math.u-strasbg.fr, boud.roukema@obspm.fr} 
}

\date{\frtoday}

\begin{document}

\maketitle

\begin{abstract}
Quel est, ou pourrait \^etre, la topologie globale de la partie spatiale
de l'Univers ? L'Univers entier (pr\'ecis\'ement, l'hypersurface spatiale
de celui-ci) est-il observable ?  Les math\'ematiciens, les physiciens
et les cosmologistes observationnels ont des approches diff\'erentes
pour aborder ces questions qui restent ouvertes.

{\em What is, or could be, the global topology of spatial sections of the
Universe? Is the entire Universe (spatial hypersurface thereof)
observable ? Mathematicians, physicists and observational cosmologists
have different strategies to approaching these questions which are not
yet fully answered.}
\end{abstract}

Un atelier international d'une journ\'ee a eu lieu \`a Paris pour les
chercheurs de ces domaines compl\'ementaires, pour introduire leurs
sujets et pour pr\'esenter des revues autant que les derniers r\'esultats
de leurs travaux. Math\'ematiciens, astronomes et physiciens y ont 
particip\'es. L'atelier a \'et\'e organis\'e dans le cadre du PNC 
(programme national de cosmologie).

Merci \`a tous les participants, \`a ceux qui ont pr\'esent\'e leur
travaux ainsi qu'\`a ceux qui ont \'ecout\'e et particip\'e aux
d\'ebats. Merci aussi \`a l'Observatoire de Paris, l'Institut
d'Astrophysique de Paris et le PNC pour leur aide.

Ce compte rendu est constitu\'e de
trois articles th\'eoriques~:
 celui de Ratcliffe et Tschantz sur un objet
math\'ematique utile pour la gravit\'e quantique, l'instanton
gravitationnel; une exploration des liens \'eventuels entre
une constante cosmologique non-nulle et la topologie cosmique
par Lachi\`eze-Rey; et e~Costa et Fagundes 
ont pr\'esent\'e un potentiel $V(\phi)$ qui pourrait
donner naissance \`a un univers multi-connexe, hyperbolique et
compact; 
et de quatres articles observationnels~: une revue par Roukema; 
un rappel par Wichoski que 
vu les difficult\'es pratiques des m\'ethodes statistiques \`a 3-D 
et \`a 2-D, 
la recherche des images topologiques de notre propre Galaxie 
ne doit pas \^etre oubli\'e;
un \'eclairci par Inoue sur le vif d\'ebat actuellement en cours 
concernant les analyses des donn\'ees du fond diffus cosmique de COBE
pour les mod\`eles hyperboliques et compactes;
et un r\'esum\'e de la m\'ethode de la reconnaissance des 
sch\'emas des taches par Levin \& Heard.

Nul peut pr\'evoir en ce moment
quels \'el\'ements th\'eoriques et observationnels seront les
plus importants, m\^eme si chacun
de nous a ses propres intuitions\ldots

Cet atelier de d\'ecembre 1998 a suivi le premier du septembre 1997
\`a Cleveland, et comme il vient d'y avoir deux s\'eances parall\`eles
sur la topologie cosmique et les 3-vari\'et\'es hyperboliques \`a la
r\`eunion Marcel Grossmann IX
\`a Roma en juillet 2000, la continuation d'un d\'eveloppement rapide 
et soutenu est promise\ldots

\newpage {\em A one-day international workshop, supported by the PNC (Programme
National de la Cosmologie), was held in Paris for members of the
different disciplines to introduce their respective subjects and
present both reviews and up-to-date research methods and
results. Mathematicians, astronomers and physicists were welcomed.

Thank you to all the participants, both to those who presented work
and those who listened and participated in the discussion. Thank you
also to the Observatoire de Paris, the Institut d'Astrophysique de
Paris, and the PNC.

In this volume, we have three theoretical articles:
that of Ratcliffe \& Tschantz about a mathematical object which 
should be useful for quantum gravity, the gravitational instanton;
an exploration of possible links between a non-zero cosmological
constant and cosmic topology by Lachi\`eze-Rey; and e~Costa \&
Fagundes presented a potential $V(\phi)$ which could give birth
to a multiply connected, compact hyperbolic universe; and four
observational articles: a review by Roukema; a reminder by Wichoski
that given the practical difficulties of statistical methods in 3-D
and in 2-D, 
the search for topological images of our own Galaxy should not be
forgotten; 
some very interesting comments by Inoue on the lively debate 
presently underway regarding compact hyperbolic model analyses of
the COBE cosmic microwave background data;
and a summary of the method of spot pattern recognition
by Levin \& Heard.

Noone can predict which theoretical and observational elements will 
be the most important, even if each of us has his or her own 
intuition\ldots

This Dec 1998 workshop followed the first in Cleveland in Sep 1997,
and as two parallel sessions on cosmic topology and hyperbolic 
3-manifolds have just taken place at the Marcel Grossmann IX meeting
in Roma in July 2000, continued rapid development and excitement 
in the field is the safest prediction to make for the near future\ldots
}

\bigskip
\noindent Comit\'e d'organisation et scientifique/
{\em Organising and Scientific Committee}: 

Boud Roukema, Vincent Blanloeil, Jean-Pierre Luminet, Gary Mamon

\bigskip
\noindent Some additional useful links are provided at the electronic site
of the proceedings at: 

{\em http://www.iap.fr/user/roukema/CTP98/programme.html}.

\bigskip
\noindent \underline{Participants:}
\medskip

{\parskip=2pt
 \noindent Bajtlik, Stanislaw, 
{\em  Nicolas Copernicus Astronomical Center, Warsaw, Poland}

 \noindent Blanl{\oe}il, Vincent, 
{\em  IRMA, Univ de Strasbourg I, France}

 \noindent Bois, Eric, 
{\em  Observatoire de Bordeaux, France}

 \noindent C\'el\'erier, Marie-No\"elle, 
{\em  DARC, Observatoire de Paris-Meudon, France}

 \noindent Fagundes, Helio, 
{\em  IFT, Univ. Estadual Paulista, Brazil }

 \noindent Gausmann, Evelise, 
{\em  Instituto de Fisica Teorica, Univ. Estadual Paulista, Brazil}

 \noindent Inoue, Kaiki Taro, 
{\em  Yukawa Institute for Theoretical Physics, Kyoto, Japan}

 \noindent Lachi\`eze-Rey, Marc, 
{\em  CEA, Saclay, France}

 \noindent Lehoucq, Roland, 
{\em  Service d'Astrophysique de Saclay, France}

 \noindent Levin, Janna, 
{\em  Astronomy Centre, Univ Sussex, United Kingdom}

 \noindent Luminet, Jean-Pierre, 
{\em  DARC, Observatoire de Paris-Meudon, France}

 \noindent Madore, John, 
{\em  Universit\'e de Paris Sud, France}

 \noindent Mamon, Gary, 
{\em  IAP, Paris, France}

 \noindent Marty, Philippe, 
{\em  Institut d'Astrophysique Spatiale, Orsay, France}

 \noindent Pierre, Marguerite, 
{\em  Service d'Astrophysique  du CE Saclay, France}

 \noindent Pogosyan, Dmitri,
{\em  CITA, Toronto, Canada}

 \noindent Ratcliffe, John G., 
{\em  Vanderbilt University, Tennessee, USA}

 \noindent Roukema, Boud, 
{\em  IUCAA, Pune, India}

 \noindent Uzan, Jean-Phillippe, 
{\em  Univ of Geneva, Switzerland}

 \noindent Van Waerbeke, Ludovic, 
{\em  CITA, Toronto, Canada}

 \noindent Weeks, Jeff, 
{\em  Canton NY, USA}

 \noindent Wichoski, Ubi, 
{\em  Brown University, USA}
}

\newpage 

\noindent \underline{Programme:}
\medskip

\noindent  Chair: Luminet

\smallskip \noindent  09h00: Jean-Pierre Luminet (DARC, Observatoire de Paris - Meudon) 
 
{\em   Cosmological Topology: Opening Remarks }


\smallskip \noindent  09h05: Jeff Weeks (Canton NY, USA) 
 
{\em   (1) Deducing topology from the CMB; (2) The structure of closed hyperbolic 3-manifolds }

\smallskip \noindent  09h55: Dmitri Pogosyan (CITA, Toronto)
 
{\em Some work on hyperbolic 3-manifolds and COBE data}

\smallskip \noindent  10h00: John Madore (Univ Paris Sud) 
 
{\em   Topology at the Planck Length }

\smallskip \noindent  10h30-11h00: coffee break

\smallskip \noindent  11h00: John G. Ratcliffe (Vanderbilt University) 
 
{\em   Gravitational Instantons of Constant Curvature }

\smallskip \noindent  11h30: Marc Lachièze-Rey (CEA, Saclay) 
 
{\em   The Physics of Cosmic Topology }

\smallskip \noindent  12h00: Boudewijn Roukema (IUCAA, Pune) 

{\em  Observational Methods, Constraints and Candidates }

\smallskip \noindent  12h30: Marguerite Pierre (Service d'Astrophysique, CEA, Saclay) 
 
{\em   X-ray Cosmic Topology }

\smallskip \noindent  13h00-14h00: Lunch 

\medskip

\smallskip \noindent Chair: Fagundes

\smallskip \noindent  14h00: Helio Fagundes (IFT, Univ Estadual Paulista) 
 
{\em   Creation of a Closed Hyperbolic Universe }

\smallskip \noindent  14h30: Ubi Wichoski (Brown Univ, USA) 
 
{\em   Topological Images of the Galaxy }

\smallskip \noindent 15h00: Jean-Phillippe Uzan (Univ of Geneva) 
 
{\em \smallskip \noindent    Cosmic Crystallography: the Hyperbolic Case }

\smallskip \noindent 15h30-16h00: coffee break

\smallskip \noindent 16h00: Kaiki Taro Inoue (Yukawa Institute for Theoretical Physics) 
 
{\em   CMB anisotropy in a compact hyperbolic universe }

\smallskip \noindent 16h30: Janna Levin (Astronomy Centre, Univ Sussex) 
 
{\em   How the Universe got its Spots }

\smallskip \noindent 17h00: Stanislaw Bajtlik (Copernicus Center, Warsaw) 

{\em  Applying Cosmo-topology: Galaxy Transverse Velocities }

\medskip

\smallskip \noindent Moderator: Roukema

\smallskip \noindent 17h30:  (all participants)

{\em General Discussion}

\smallskip \noindent 18h00: Close

\newpage \tableofcontents \listoffigures \listoftables


\newtheorem{theorem}{Theorem}
\newtheorem{lemma}[theorem]{Lemma}
\newtheorem{corollary}{Corollary}


\input amssym1997.sty
\newcommand{\integers}{\mbox{$Z$\hskip -1ex$Z$}}

\newcommand{\realnos}{{\mathbb R}}
\newcommand{\rationalnos}{{\mathbb Q}}
\newcommand{\complexnos}{{\mathbb C}}
\newcommand{\p}{\phantom{-}}
\font\eighttt=cmtt8
\font\tentt=cmtt10

\subtitle{Gravitational Instantons of Constant Curvature}

\subauthor{John G. Ratcliffe, Steven T. Tschantz}
\subaddress{Department of Mathematics, Vanderbilt University,\\
 Nashville, Tennessee 37240, U.S.A. 
}

\date{}


\submaketitle


\begin{abstract}
In this paper, we classify all closed flat 4-manifolds 
that have a reflective symmetry along a separating totally geodesic hypersurface.
We also give examples of small volume hyperbolic 4-manifolds 
that have a reflective symmetry along a separating totally geodesic hypersurface.
Our examples are constructed by gluing together polytopes in hyperbolic 4-space. 
\end{abstract}

\subsection{Introduction}

In a recent paper [3], G.W. Gibbons mentioned that the 
examples of minimum volume hyperbolic 4-manifolds described in our paper [7] 
might have applications in cosmology. 
In this paper, we elaborate on our examples and introduce some new examples. 
In particular, we construct an example which answers in the affirmative 
the following question posed by G.W. Gibbons 
at the Cleveland Cosmology-Topology Workshop.  
Can one find a closed hyperbolic 4-manifold with 
a (connected) totally geodesic hypersurface that separates?  
In this paper a {\it hypersurface} is a codimension one submanifold.  
We begin by describing the geometric setup of real tunneling geometries. 

According to Gibbons [3], current models of the quantum origin of the universe 
begin with a {\it real tunneling geometry}, that is, a solution of the classical 
Einstein equations which consists of a Riemannian 4-manifold $M_R$ 
and a Lorentzian 4-manifold $M_L$ joined across a totally geodesic 
spacelike hypersurface $\Sigma$ which serves as an initial Cauchy 
surface for the Lorentzian spacetime $M_L$. In cosmology, $\Sigma$ 
is taken to be closed, that is, compact without boundary, 
and in accordance with the No Boundary Proposal  
one usually takes $M_R$ to be connected, 
orientable, and compact with boundary equal to $\Sigma$. 

Given this setup one may pass to the double $2M_R=M_R^+ \cup M_R^-$ 
by joining two copies of $M_R$ across $\Sigma$.  This is a closed 
orientable Riemannian 4-manifold $M = 2M_R$ called the {\it gravitational instanton} 
of the real tunneling geometry. 
The instanton $M$ admits a reflection map $\theta$ 
that is an orientation reversing involution which fixes the totally 
geodesic submanifold $\Sigma$ and permutes the two portions $M_R^\pm$. 
According to Gibbons, the involution $\theta$ plays a crucial role 
in the quantum theory because it allows one to formulate the requirement 
of ``reflection positivity". 

In the standard example of a real tunneling geometry, 
the instanton $M$ is the unit 4-sphere $S^4$ 
and $\Sigma$ is the unit 3-sphere $S^3$ 
thought of as the equator of $S^4$.
The 3-sphere $S^3$ is the simplest model of the universe 
that is isotropic and the 4-sphere $S^4$ is 
the only gravitational instanton that is isotropic. 

The Riemannian manifolds that are locally isotropic 
are the manifolds of constant sectional curvature $k$. 
For simplicity, a Riemannian manifold of constant sectional 
curvature is usually normalized to have curvature $k=-1,0$, or 1. 
A Riemannian manifold of constant sectional curvature $k=-1,0$, or 1 
is called a {\it hyperbolic, Euclidean}, or {\it spherical} 
manifold, respectively. Euclidean manifolds are also called {\it flat} manifolds.  
We shall assume that a hyperbolic, Euclidean, or spherical manifold is 
connected and complete. We shall also assume that a manifold does not 
have a boundary unless otherwise stated. 
Then a hyperbolic, Euclidean, or spherical $n$-manifold is isometric to 
the orbit space $X/\Gamma$ of a freely acting discrete group of isometries $\Gamma$ 
of hyperbolic, Euclidean, or spherical $n$-space $X = H^n, E^n$, or $S^n$, 
respectively. A discrete group $\Gamma$ of isometries of $H^n$ or $E^n$ acts 
freely on $H^n$ or $E^n$ if and only if $\Gamma$ is torsion-free. 

\subsection{Spherical and Flat Gravitational Instantons}  

The first observation to make about real tunneling geometries is 
that the Euler characteristic of a gravitational instanton $M$ is even, since 
$$\chi(M) = \chi(M_R^+) + \chi(M_R^-) - \chi(\Sigma) = 2\chi(M_R^+).$$
There are only two spherical 4-manifolds, namely $S^4$ 
and elliptic $4$-space $P^4$ (real projective 4-space). 
Spherical $4$-space $S^4$ is the prototype for a gravitational instanton 
whereas $P^4$ is not a gravitational instanton, since $\chi(P^4) = 1$.
 
We next classify Euclidean (or flat) gravitational instantons. 
In order to state our classification, we need to recall the definition 
of a twisted $I$-bundle. Let $N$ be a nonorientable $n$-manifold. 
Then $N$ has an orientable double cover $\tilde N$ 
and there is a fixed point free, orientation reversing involution 
$\sigma$ of $\tilde N$ such that $N$ is the quotient 
space of $\tilde N$ obtained by identifying $\sigma(x)$ with $x$ 
for each point $x$ of $\tilde N$. Let $I$ be a closed interval $[-b,b]$ with $b>0$. 
Then $\sigma$ extends to a fixed point free, orientation preserving involution 
$\tau$ of $\tilde N \times I$ defined by $\tau(x,t) = (\sigma(x),-t)$. 
The {\it twisted I-bundle} $B$ over $N$ is the quotient space of $\tilde N \times I$ 
obtained by identifying $\tau(x,t)$ with $(x,t)$ for each point $(x,t)$ 
of $\tilde N \times I$. Then $B$ is an orientable $(n+1)$-manifold 
with boundary $\partial B=\tilde N$ and $\tilde N \times I$ is a double 
cover of $B$.  Note that $B$ is a fiber bundle over $N$ with fiber $I$. 
If $N$ is a Riemannian manifold, then $\tilde N$ has a Riemannian metric 
so that the double covering from $\tilde N$ to $N$ is a local isometry. 
We give $I$ the standard Euclidean metric and $\tilde N \times I$ 
the product Riemannian metric. Then $\tau$ is an isometry 
of $\tilde N \times I$ and so $B$ inherits a Riemannian metric, 
which we called the {\it twisted product Riemannian metric}, 
so that the double covering from $\tilde N \times I$ to $B$ is a local isometry.  
It is worth mentioning that twisted $I$-bundles occur naturally in topology, 
since a closed regular neighborhood of a nonorientable hypersurface $N$ 
of an orientable manifold $M$ is a twisted $I$-bundle over $N$. 
We are now ready to state our classification theorem for flat 
gravitational instantons. We shall state our theorem 
in arbitrary dimensions, since the proof works in all dimensions. 

\begin{theorem} 
Let $M$ be a connected, closed, orientable, Riemannian $n$-manifold 
that is obtained by doubling a Riemannian $n$-manifold $M_R$ with 
a totally geodesic boundary $\Sigma$. Then 
(1)  $M$ is flat and $\Sigma$ is connected if and only if 
$M_R$ is a twisted $I$-bundle, with the twisted product 
Riemannian metric, over a connected, closed, nonorientable, 
flat $(n-1)$-manifold $N$; and, 
(2) $M$ is flat and $\Sigma$ is disconnected if and only if 
$M_R$ is a product $I$-bundle, $N\times I$ with the product Riemannian metric, 
over a connected, closed, orientable, flat $(n-1)$-manifold $N$.  
\end{theorem}

{\em Proof.}
Assume that $M$ is flat. Then $M$ is complete, since $M$ is compact. 
Hence we may assume  $M = E^n/\Gamma$ where $\Gamma$ is a freely acting  
discrete group of orientation preserving isometries of Euclidean $n$-space $E^n$. 
Let $\phi:E^n \to E^n/\Gamma$ be the quotient map. 
Then $\phi$ is a universal covering projection. 
Now $\phi^{-1}(\Sigma)$ is a totally geodesic hypersurface of $E^n$, 
since $\phi$ is a local isometry. 
Therefore $\phi^{-1}(\Sigma)$ is a disjoint union of hyperplanes of $E^n$. 

The inclusion map $\iota:\Sigma \to M$ induces an injection 
$\iota_\ast:\pi_1(\Sigma)\to\pi_1(M)$ on fundamental groups, 
since each component of $\phi^{-1}(\Sigma)$ is simply connected. 
Choose a component $P$ of $\phi^{-1}(\Sigma)$. 
We may identify $\pi_1(M)$ with $\Gamma$ and $\pi_1(\Sigma)$ 
with the subgroup $C$ of $\Gamma$ that leaves $P$ invariant. 
The group $\Gamma$ has cohomological dimension $n$ 
and the group $C$ has cohomological dimension $n-1$, 
since $M$ and $\Sigma$ are aspherical manifolds. 
Now every subgroup of finite index of an $n$-dimensional group 
is $n$-dimensional. Therefore the index of $C$ in $\Gamma$ is infinite. 
The number of components of $\phi^{-1}(\Sigma)$ is the index of $C$ in $\Gamma$. 
Therefore $\phi^{-1}(\Sigma)$ is a disjoint union of an infinite number of 
hyperplanes of $E^n$. These hyperplanes are all parallel, 
since any two nonparallel hyperplanes of $E^n$ intersect. 

Color $M_R$ white and the rest of $M$ black. 
The boundary between the white and black regions of $M$ is $\Sigma$. 
Lift this coloring to $E^n$ via $\phi: E^n\to M$ 
by coloring $\phi^{-1}(M_R)$ white and the rest of $E^n$ black. 
Then the regions between the hyperplanes of $\phi^{-1}(\Sigma)$ 
are colored alternately white and black, 
since the coloring must change at each hyperplane of $\phi^{-1}(\Sigma)$. 
Let $R$ be the component of $\phi^{-1}(M_R)$ containing $P$. 
Then $R$ is the closed white region bounded by two adjacent hyperplanes 
$P$ and $Q$ of $\phi^{-1}(\Sigma)$. 
Hence $R$ is a Cartesian product $P\times I$ where $I$ is a closed 
line segment running perpendicularly from $P$ to $Q$. 
Therefore $R$ is simply connected, and so the inclusion map $\kappa:M_R\to M$ 
induces an injection $\kappa_\ast:\pi_1(M_R)\to\pi_1(M)$ on fundamental groups. 
Hence we may identify $\pi_1(M_R)$ with the subgroup $A$ of $\Gamma$ 
that leaves $R$ invariant.

Let $H$ be the hyperplane of $E^n$ midway between $P$ and $Q$. 
Then $H$ cuts $I$ at its midpoint. Now each element of $A$ maps 
$I$ to a line segment running perpendicularly from $P$ to $Q$, 
since each element of $A$ leaves $\partial R = P\cup Q$ invariant and 
preserves perpendicularity. Therefore $A$ leaves $H$ invariant. 
Hence $H/A$ is a hypersurface of $M_R = R/A$. 

Assume first that $\Sigma$ is disconnected. 
Then no element of $A$ interchanges $P$ and $Q$, 
and so $A$ leaves both $P$ and $Q$ invariant. 
Hence $A$ preserves the product structure $R = H \times I$. 
Therefore $R/A = H/A \times I$.  Now since $A$ preserves the orientation of $E^n$ 
and preserves both sides of $H$ in $R$, 
we deduce that $A$ preserves orientation on $H$. 
Therefore $N = H/A$ is an orientable manifold and $M_R$ 
is the product $I$-bundle, $N \times I$, with the product Riemannian metric. 
Moreover, $N$ is a closed manifold, since $M_R$ and $N$ are compact. 

Now assume that $\Sigma$ is connected. 
Then there is an element $\alpha$ of $A$ that interchanges 
$P$ and $Q$.  Let $B$ be the subgroup of $A$ 
that leaves both $P$ and $Q$ invariant. 
Then $B$ is a subgroup of $A$ of index two. 
Now $B$ preserves the product structure $R = H\times I$. 
Therefore $R/B = H/B\times I$. 
Now since $A$ preserves the orientation of $E^n$, 
with $B$ preserving both sides of $H$ in $R$ 
and $\alpha$ interchanging both sides of $H$ in $R$, 
we deduce that $B$ preserves orientation on $H$ 
and $\alpha$ reverses orientation on $H$. 
Therefore $\tilde N = H/B$ is the orientable double cover of the 
nonorientable manifold $N = H/A$. 
Now $A/B$ acts on $\tilde N \times I$ so that 
$\tilde N\times I/(A/B) = R/A$ is a twisted 
$I$-bundle over $H/A$. 
Thus $M_R$ is a twisted $I$-bundle, with the twisted product Riemannian metric, 
over the nonorientable manifold $N$. 
Now $M_R$ is compact and so its double cover $\tilde N \times I$ is compact. 
Hence $\tilde N$ and $N$ are compact, and so $N$ is a closed manifold. 

Conversely, if $M_R$ is an $I$-bundle over a flat $(n-1)$-manifold $N$,  
with either the product or twisted product Riemannian metric, 
and a totally geodesic boundary, then obviously $M = 2M_R$ is flat. 
\hfill \fbox{\rule{0ex}{1ex}\rule{1ex}{0ex}}
\bigskip

There are exactly 10 closed flat 3-manifolds up to affine equivalence. 
Six of these manifolds are orientable and four are nonorientable. 
We shall denote the orientable manifolds by $O_1, O_2,\ldots, O_6$ 
and the nonorientable manifolds by $N_1,N_2,N_3,N_4$. 
As a reference for closed flat 3-manifolds, see Wolf [8]. 
We shall take the same ordering of the closed flat 3-manifolds as in Wolf [8]. 
In particular, the 3-manifold $O_1$ is a flat 3-torus. 

Let $M$ be a flat gravitational instanton. 
Then $M$ is a connected, closed, orientable, flat 4-manifold that is obtained 
by doubling a flat Riemannian 4-manifold $M_R$ with totally 
geodesic boundary $\Sigma$. 
Assume first that $\Sigma$ is disconnected.  
Then $M_R$ is a product $I$-bundle $O \times I$, 
with the product Riemannian metric, over a closed orientable flat 
3-manifold $O$ by Theorem 1.  This implies that $M_R$ is just a straight tube 
with opening and closing end isometric to $O$. 
Here $\Sigma = \partial M_R$ is the disjoint union of two isometric copies of $O$. 
One can interpret the geometry of $M_R$ as leading to the birth 
of disjoint identical twin Lorentzian universes or, by reversing the arrow of time 
in one of the universes, as a collapse and subsequent rebirth 
of a Lorentzian universe.  

\begin{figure}
\centerline{\epsfig{file=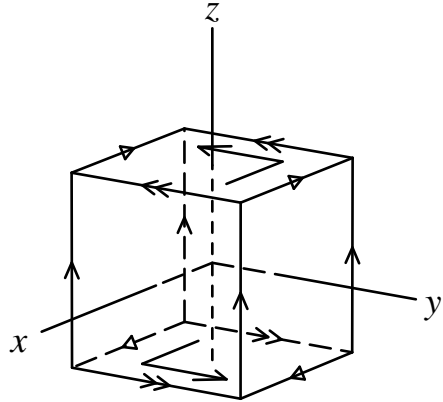}}
\vspace{-.4in}
\caption{A fundamental domain for the half-twisted 3-torus}
\label{f-rat1}
\end{figure}

Assume now that $\Sigma$ is connected.  
Then $M_R$ is a twisted $I$-bundle, 
with the twisted product Riemannian metric, over a closed 
nonorientable flat 3-manifold $N$ by Theorem 1. 
Here $\Sigma$ is the orientable double cover of $N$. 
According to Theorem 3.5.9 of Wolf [8], 
if $N$ is affinely equivalent to $N_1$ or $N_2$, 
then $\Sigma$ is a flat 3-torus, whereas  
if $N$ is affinely equivalent to $N_3$ or $N_4$, 
then $\Sigma$ is affinely equivalent to $O_2$.
Thus only the first two affine equivalence types of closed orientable 
flat 3-manifolds are possible initial hypersurfaces for the creation 
of a connected Lorentzian universe from a flat gravitational instanton.

We call the closed orientable flat 3-manifold $O_2$ a {\it half-twisted 
3-torus} because $O_2$ can be constructed from a rectangular box, 
centered at the origin in $E^3$ with sides parallel to the coordinate planes, 
by identifying opposite pairs of vertical sides by translations and 
identifying the top and bottom sides by a half-twist in the $z$-axis. 
See Figure~\ref{f-rat1}.  
The first homology group of $O_2$ is $\integers\oplus\integers_2\oplus\integers_2$. 
Therefore $O_2$ is not topologically equivalent to a 3-torus. 
It is worth noting that $O_2$ is double covered by a 3-torus. 
This is easy to see by stacking two of the boxes defining $O_2$ on top 
of each other.

\subsection{Hyperbolic Gravitational Instantons}

A {\it hyperbolic gravitational instanton} is a gravitational  
instanton that is a hyperbolic manifold. 
Thus a hyperbolic gravitational instanton is a closed, orientable
hyperbolic 4-manifold, $M$, with a separating, totally geodesic,
orientable, hypersurface $\Sigma$ which is the set of
fixed points of an orientation reversing isometric involution of $M$. 
As a reference for hyperbolic manifolds, see Ratcliffe [6]. 
Cosmologists are interested in small volume hyperbolic gravitational instantons 
because the probability of creation of a hyperbolic gravitational instanton  
increases with decreasing volume.   
The volume of a hyperbolic 4-manifold $M$ of finite volume is proportional 
to its Euler characteristic $\chi(M)$ and so the Euler characteristic 
is an effective measure of the volume of a hyperbolic gravitational instanton. 
The closed orientable hyperbolic 4-manifold of least known volume is 
the Davis hyperbolic 4-manifold [2], which has Euler characteristic 26. 

In his talk at the Cleveland Cosmology-Topology Workshop,  G.W. Gibbons 
asked the question: 

\begin{quote}
\em Can one find a closed hyperbolic 4-manifold with 
a totally geodesic two-sided hypersurface that separates? 
\end{quote} 
 
\noindent 
It is well known that there are closed hyperbolic 4-manifolds with 
two-sided totally geodesic hypersurfaces.  As pointed out by Gibbons [3], 
if a two-sided hypersurface $\Sigma$ of a manifold $M$ does not separate, 
then $M$ has a double cover with a separating hypersurface 
consisting of two disjoint copies of $\Sigma$.  
Thus an affirmative answer to Gibbon's question  
has been known for some time with $\Sigma$ disconnected. 
See for example, \S 2.8.C of [4]. 
However, in Gibbon's paper [3], he asks whether the creation 
of a {\bf single} universe is possible from a hyperbolic gravitational instanton.  
Thus a more interesting question (and probably what Gibbons really 
wanted to ask at the workshop) is the question:  

\begin{quote}
\em Can one find a closed hyperbolic 4-manifold with 
a {\bf connected} totally geodesic two-sided hypersurface that separates? 
\end{quote} 

We will answer this question in the affirmative by constructing 
a hyperbolic gravitational instanton $M$ 
with a connected initial hypersurface $\Sigma$. 
The manifold $M$ is most easily understood as
the orientable double cover of a manifold specified by a side-pairing
of the same regular hyperbolic polytope as that used in the
construction of the Davis hyperbolic 4-manifold [2], 
and so we consider the construction of this 
manifold first.

A {\it regular {\rm 120}-cell} is a 4-dimensional, regular, convex polytope with 120
sides, each a regular dodecahedron.  Each side meets its
twelve neighbors along a pentagonal ridge (2-dimensional face).  
Each edge of the 120-cell is shared by three sides, and each vertex is shared by
four sides.  There are a total of 720 ridges, 1200 edges, and 600 vertices
in a regular 120-cell.  As the edge length of a regular hyperbolic
120-cell is increased, the dihedral angle between adjacent sides decreases.
Regular hyperbolic 120-cells with dihedral angles  
of $2\pi/3$, $\pi/2$, and $2\pi/5$ are possible 
and each can be used to tessellate hyperbolic
4-space with 3, 4, or 5 of the 120-cells fitted around each ridge respectively.  
The set of isometries of hyperbolic 4-space preserving
one of these tessellations will be a discrete group; 
the quotient of hyperbolic 4-space under the action
of a torsion-free subgroup of finite index in this group will give
a closed hyperbolic 4-manifold which can be realized by gluing together 
some number of copies of the corresponding regular 120-cell.
The Euler characteristic of the hyperbolic orbifold determined 
by a regular 120-cell, with dihedral angle
$2\pi/3$, $\pi/2$, and $2\pi/5$ is $1$, $17/2$, and $26$, respectively; 
their volumes are proportional to their Euler characteristic.

A purely combinatorial search for manifolds based on
gluing one or two of the 120-cells with dihedral angle $2\pi/3$
is essentially intractable.  Searches for side-pairings meeting some
simple restrictions have failed to uncover small volume hyperbolic
4-manifolds based on this smallest regular 120-cell.  A manifold
based on the 120-cell with dihedral angle $\pi/2$ can only result from
a gluing of an even number of 120-cells.  In fact, we have constructed
two different manifolds by gluing just two right-angled 120-cells.
These have Euler characteristic 17, are nonorientable, and do not 
seem to have the kind of totally geodesic hypersurfaces desired.

Let $P$ be a regular hyperbolic 120-cell with dihedral angles $2\pi/5$.
For simplicity, realize $P$ in the conformal ball model of hyperbolic
4-space with center at the origin and aligned so the center of a side
lies along each of the coordinate axes, i.e., there are centers of sides
having coordinates $(x_1,x_2,x_3,x_4)=(\pm r,0,0,0)$, $(0,\pm r,0,0)$,
$(0,0,\pm r,0)$, and $(0,0,0,\pm r)$ for an appropriate $r$. 
Then the four coordinate hyperplanes of $E^4$, 
given by $x_i = 0$, for $i = 1,2,3,4$, are planes of symmetry of $P$. 
A side-pairing map for $P$ can be described as a symmetry of $P$
taking a side $S$ to another side $S'$ followed by reflection in the side $S'$.  
Thus side-pairing maps will be determined by the orthogonal transformations of $E^4$ 
that are symmetries of $P$.  

The side of $P$ lying along the positive $x_4$-axis will be referred to as
the {\it side at the north pole}, the side on the negative $x_4$-axis will
be referred to as the {\it side at the south pole}, and the hyperplane with $x_4=0$,
will be referred to as the {\it equatorial plane} of $P$.  
There are 30 sides of $P$ centered on the equatorial plane and 12 ridges lie
entirely in this hyperplane.  The intersection of the equatorial plane
with $P$ is a truncated, hyperbolic, ultra-ideal triacontahedron. 

A {\it triacontahedron} is a quasiregular convex polyhedron 
with 30 congruent rhombic sides. As a reference for the geometry 
of a triacontahedron, see Coxeter [1]. 
In a triacontahedron five rhombi meet at each vertex with acute angles 
and three rhombi meet at each vertex with obtuse angles. 
A {\it hyperbolic ultra-ideal triacontahedron} is a triacontahedron 
centered at the origin in the projective disk model of hyperbolic 3-space 
whose order 5 vertices lie outside the model (hence are ultra-ideal)  
and whose order 3 vertices lie inside the model.  
A {\it truncated ultra-ideal triacontahedron} is obtained from 
an ultra-ideal triacontahedron by truncating its order 5 vertices 
yielding a polyhedron with 12 pentagonal sides corresponding to 
the order 5 vertices and 30 hexagonal sides corresponding to the 
truncated 30 rhombic sides of the triacontahedron. 

The points of $P$ with $x_4>0$ will be referred to 
as the {\it northern hemisphere} of $P$ 
while the points of $P$ with $x_4<0$ will be referred to 
as the {\it southern hemisphere} of $P$.
There are thus 45 sides of $P$ centered in the northern hemisphere: the side
at the north pole, the 12 sides adjacent to that side, 12 sides sitting
on the equatorial plane (i.e., having a ridge lying in the
equatorial plane), and 20 other sides, symmetrically positioned with
centers having the same $x_4$-coordinate, that fill in the gaps between the
two layers of 12 and the sides centered on the equatorial plane.

The Davis hyperbolic 4-manifold $M_0$ is realized as a gluing of 
the 120-cell $P$ defined by the following side-pairing maps.  
For each side $S$ of $P$, take $S'$ to be the antipodal side of $P$, 
and let the side-pairing map from $S$ to $S'$ be reflection in the hyperplane 
which is the perpendicular bisector of the line segment between the centers
of $S$ and $S'$, followed by reflection in side $S'$.
Thus, for example, the side at the north pole is reflected in the equatorial
plane to the side at the south pole.  Each side centered on the
equatorial plane is paired to another side centered on the equatorial plane
so that points of that side in the northern hemisphere map to points
of the other side also in the northern hemisphere.  Each side
centered in the northern hemisphere is side-paired with one centered
in the southern hemisphere.  
To see that this gluing results in a hyperbolic 4-manifold, 
it is necessary to check that the ridges are identified in cycles of 5, 
and that the edges and vertices of $P$ are similarly
identified so that the correct number of each belong to a cycle and
a solid ball is formed around each edge and vertex equivalence class
in the manifold, in this case, there must be 20 edges in each edge cycle
and all 600 vertices of $P$ must form a single vertex cycle.

\begin{figure}
\centerline{\epsfig{file=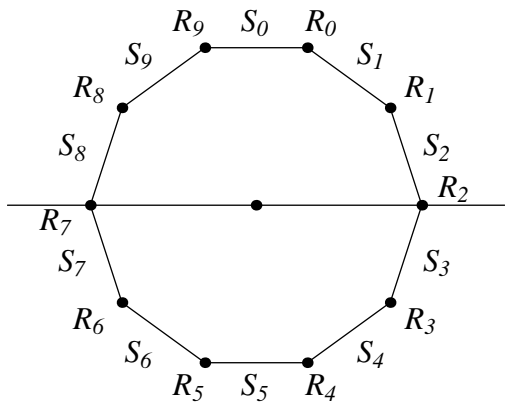}}
\caption{A chain of adjacent sides of a regular 120-cell}
\label{f-rat2}
\end{figure}

Suppose $S_0$ is any side of $P$, and $R_0$ is a ridge of $S_0$.
Let $S_1$ be the side adjacent to $S_0$ along $R_0$, and let $R_1$
be the ridge opposite $R_0$ in the side $S_1$.  Continue in this
manner taking $S_{i+1}$ adjacent to $S_i$ along $R_i$ and $R_{i+1}$
opposite $R_i$ in $S_{i+1}$.  Then $S_{10}=S_0$ and $R_{10}=R_0$. 
For example, if $S_0$ is the side at the north pole, $R_0$ is a ridge
of $S_0$, then $S_1$ is one of the twelve immediate neighbors to $S_0$.
The side $S_2$ adjacent to $S_1$ along $R_1$ is one of the twelve northern
hemisphere sides sitting on the equatorial plane and $R_2$ is its ridge
in the equatorial plane.  Sides $S_3$, $S_4$, $S_5$, $S_6$, and $S_7$
are in the southern hemisphere with $S_5$ the side at the south pole
and $S_7$ adjacent to $S_8$ along the ridge $R_7$ antipodal to $R_2$
in the equatorial plane.  Finally, side $S_8$ and $S_9$ are back in
the northern hemisphere with $S_9$ the side adjacent to $S_0$
along the ridge $R_9$ opposite the original $R_0$.  
See Figure~\ref{f-rat2}. 
The ridge $R_0$ of $S_0$ is identified with $R_4$ of $S_5$ by the side-pairing map 
of the side at the north pole with the side at the south pole.
In turn, $R_4$ is identified with $R_8$ by the side-pairing map
of $S_4$ to $S_9$, which is identified with $R_2$ by the map
of $S_8$ to $S_3$, and then identified with $R_6$ by the map
of $S_2$ to $S_7$, and back to $R_0$ by the map of $S_6$ to $S_1$.
Thus each ridge cycle consists of 5 ridges of $P$.  The edge and
vertex cycles can also be checked.

\begin{figure}
\begin{centering}
\centerline{\epsfig{file=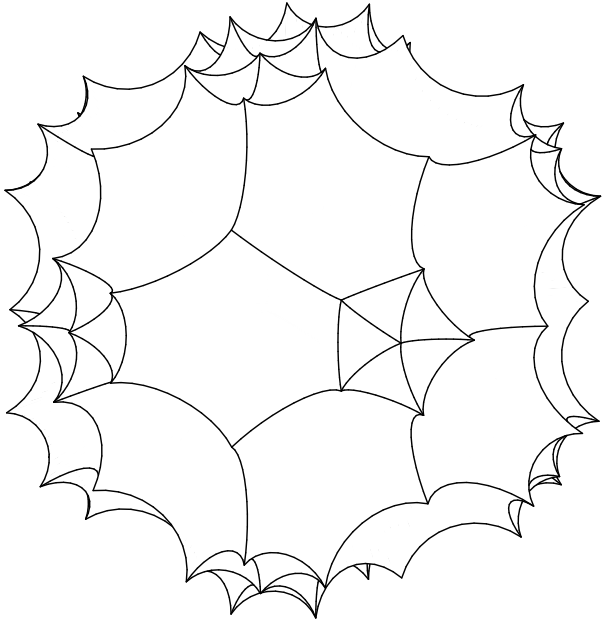}}
\end{centering}
\caption{A fundamental domain for the Davis manifold cross-section}
\label{f-rat3}
\end{figure}

Also of significance in this analysis of ridge cycles is that a ridge
$R_0$ of the side $S_0$ at the north pole, and the corresponding ridge
$R_4$ of the side $S_5$ at the south pole, are identified (in two steps)
with a ridge $R_2$ in the equatorial plane.  Consideration of the link of
this ridge in the glued-up manifold leads to the conclusion that the
equatorial cross-section of $P$ extends geodesically in the manifold to include
the identified sides at the north and south poles.  Here it is
useful to consider the gluing of a hyperbolic regular decagon with
dihedral angles $2\pi/5$ defined similarly by reflecting one side
to its antipodal side in the perpendicular bisector of the line segment 
joining their centers.  
One can more easily see how the line
connecting opposite vertices of the decagon extends to include
identified sides in the resulting glued-up 2-manifold.  
Thus the Davis hyperbolic 4-manifold $M_0$ contains, 
as a totally geodesic hypersurface $\Sigma_0$,
the equatorial cross-section of $P$ together with the identified sides at
the north and south poles.  If we subdivide these identified dodecahedra
by taking 12 pentagonal cones from each ridge to the center of
the dodecahedron and attach these cones to the pentagonal sides  
of the truncated ultra-ideal triacontahedron equatorial cross-section of $P$,  
we get the polyhedron pictured in Figure~\ref{f-rat3}.  The hexagons meet each other at
angles of $2\pi/5$, the hexagons meet triangles at angles of $\pi/2$,
and the triangles meet each other at angles of $2\pi/3$.  
The totally geodesic hypersurface $\Sigma_0$ of $M_0$ 
is obtained from this polyhedron by identifying each hexagon with
its antipodal hexagon by reflecting in the plane which is the perpendicular 
bisector of the line segment between their centers, 
and identifying each triangle with the triangle with 
which it shares a common hexagonal neighbor by a reflection in a plane perpendicular 
to that common hexagonal side. 
The homology groups of the Davis hyperbolic 4-manifold $M_0$ are 
$H_0(M_0) = \integers$, 
$H_1(M_0)= \integers^{24}$,
$H_2(M_0)= \integers^{72}$, 
$H_3(M_0)= \integers^{24}$, and 
$H_4(M_0)= \integers$. 
The homology groups of the cross-section $\Sigma_0$ of $M_0$ are 
$H_0(\Sigma_0)=\integers$, 
$H_1(\Sigma_0)=\integers^{16}$, 
$H_2(\Sigma_0)=\integers^{16}$, and 
$H_3(\Sigma_0)=\integers$.

The Davis hyperbolic 4-manifold is a closed orientable 4-manifold $M_0$ 
having a totally geodesic orientable hypersurface $\Sigma_0$ which is
a mirror for $M_0$; however, $\Sigma_0$ does not separate $M_0$, 
since we have side-pairing maps that go from the northern
hemisphere to the southern hemisphere.  To repair this last difficulty
we modify the Davis manifold side-pairing.
If $S$ is the side at the north or south pole or a side centered in
the equatorial plane we take the same side-pairing map of $S$ to its
antipodal side $S'$.  Otherwise consider the side-pairing of $S$
to the side $S'$ which is the composition of the reflection
in the equatorial plane with the side-pairing map used in the Davis
manifold, i.e., $S'$ is taken to be the reflection in the
equatorial plane of the side $S''$ antipodal to $S$ and
the side-pairing map of $S$ to $S'$ is the composition
of reflection of $S$ to $S''$ in the hyperplane which is the perpendicular bisector 
of the line segment between their centers, the reflection in the equatorial plane
taking $S''$ to $S'$, followed by, as usual, the reflection in side $S'$.
Points in the northern hemisphere are thus identified with points
in the northern hemisphere except that the side at the north pole
is identified with the side at the south pole.

Consider then how the ridge cycles in this gluing correspond to ridge
cycles in the Davis manifold gluing.  A ridge cycle of the Davis manifold gluing 
including a ridge centered on, but not contained in, the equatorial plane
is left unchanged since all of the side-pairings involving such ridges
are of sides centered on the equatorial plane and none of these side-pairings 
have been changed.  A ridge cycle of the Davis manifold gluing 
not including a ridge centered on the equatorial plane involves three
ridges on one side of the equatorial plane and two on the other.
Such a ridge cycle will not involve a ridge of the sides at the
north or south poles since these ridge cycles include also a ridge
in the equatorial plane.  The side-pairings for such a ridge cycle
will include just one side-pairing between sides centered on the
equatorial plane.  In the new side-pairing, the corresponding
ridge cycles will result from adding an extra reflection
in the equatorial plane to the side-pairings that
cross from one hemisphere to the other, that is, the ridge cycle of a
ridge in the northern hemisphere is obtained by taking the ridge cycle
in the Davis manifold gluing and reflecting those ridges that lie
in the southern hemisphere back into the northern hemisphere.
For the ridge cycle of a ridge $R_0$ of the side $S_0$ at the north pole
we get $R_0$ identified with $R_4$ by the map of $S_0$ to $S_5$,
then identified with $R_6$ by the map of $S_4$ to $S_6$ (reflected from
$R_8$ in $S_9$), identified with $R_2$ by the map of $S_7$ to $S_3$
(reflected from $R_2$ in $S_2$), and then, in the northern hemisphere,
identified with $R_8$ by the map of $S_2$ to $S_8$, and back to $R_0$
by the map of $S_9$ to $S_1$.  
See Figure~\ref{f-rat2}. 
The edge cycles and vertex cycles can also be checked 
and the side-pairing thus defines a gluing of $P$
resulting in a hyperbolic 4-manifold $M_1$.

Consideration of the ridge cycle in $M_1$ of a ridge contained in the
equatorial plane of $P$ leads to the conclusion that the equatorial
cross-section of $P$ extends geodesically in $M_1$ to include the identified sides
at the north and south poles in exactly the same way as it does in
the Davis manifold.  The conclusion is that $M_1$ contains,
as a totally geodesic hypersurface $\Sigma_1$, the same cross-section as we had in
the Davis manifold.  This hypersurface $\Sigma_1 = \Sigma_0$ is now separating,  
since the equatorial cross-section 
and the identified sides at the north and south poles
separate the northern hemisphere from the southern hemisphere in the
glued-up manifold.  The hypersurface $\Sigma_1$ is also a mirror for $M_1$.
The existence of the manifold $M_1$ answers in the affirmative 
Gibbon's question; however, $M_1$ is  nonorientable
since we have added an extra reflection to the side-pairing maps that crossed
between hemispheres.  The orientable double cover of $M_1$ is
a compact, orientable, 4-manifold having two copies of $\Sigma_1$, 
since $\Sigma_1$ is orientable, which together are separating, 
totally geodesic, and a mirror for the double cover.

If we want an orientable double cover of a nonorientable 4-manifold
with separating totally geodesic hypersurface to have
a connected, separating, totally geodesic hypersurface, we need 
the hypersurface of the nonorientable 4-manifold to also be nonorientable.
A further modification of the side-pairing for $M_1$ will do the trick.
Consider the hyperplane with $x_3=0$.  It is perpendicular to the equatorial
plane and has intersection with $P$ congruent to the intersection
of the equatorial plane with $P$.  Proceed to modify the side-pairing 
for $M_1$ in the same manner as the modification
to the side-pairing of the Davis manifold,
only now with respect to this polar hyperplane.  We note that each
side centered on the hyperplane $x_3=0$ is paired in the side-pairing defining $M_1$
with another side centered on this hyperplane and we leave such side-pairings
unchanged.  The sides centered on the $x_3$-axis are in the equatorial
plane and we leave their pairing in $M_1$ unchanged.  
Every other side is paired with a side in the opposite hemisphere 
with respect to the hyperplane $x_3=0$.  
If $S$ is such a side and was paired with $S''$ in the side-pairing 
defining $M_1$, then $S$ will be paired instead with
the side $S'$ which is the reflection in the hyperplane $x_3=0$ of the side $S''$
and the side-pairing map of $S$ will be the orthogonal map pairing $S$ to $S''$  
composed with reflection in the hyperplane $x_3=0$, followed by reflection in $S'$.  
Note that the side at the north pole
is in the hyperplane $x_3=0$ and so is still paired with the side at the
south pole.  Otherwise, if $S$ is in the northern hemisphere, then it is
paired to an $S'$ also in the northern hemisphere.  The sides centered on
the equatorial plane are still paired to sides centered on the equatorial
plane, the parts in the northern hemispheres being identified.
Again we can verify ridge cycles contain 5 ridges, the ridges in
a ridge cycle of the original Davis manifold gluing are replaced by ridges
that are reflected in one or both of the equatorial plane and the hyperplane
$x_3=0$.  Edge and vertex cycles can also be verified so that the
defined side-pairing gives rise to a hyperbolic 4-manifold $M_2$.

The ridge cycles of ridges in the equatorial plane are still 
such that the geodesic extension of the equatorial cross-section of $P$ in $M_2$
includes the identified sides at the north and south poles 
and this hypersurface $\Sigma_2$ separates $M_2$ into two components.  
The hypersurface $\Sigma_2$ can be obtained
from the same polyhedron in Figure~\ref{f-rat3} by the same gluing of triangles
but a modification of the gluing of hexagons that are not centered in the
hyperplane $x_3=0$ or centered along the $x_3$-axis by reflecting in the
hyperplane $x_3=0$.  The manifolds $M_2$ and $\Sigma_2$ are nonorientable.  
Let $M$ be the orientable double cover of $M_2$. 
Then $\Sigma_2$ lifts to a connected, separating, totally geodesic, orientable 
hypersurface $\Sigma$ of $M$ which is, in fact, a mirror for $M$.  
It should be noted that the hyperplane $x_3=0$ also extends in $M_2$ 
to a separating, totally geodesic hypersurface of $M_2$, 
but it is isometric to $\Sigma_2$, 
since the construction of $M_2$ could just as well be described
by first reflecting side-pairing maps of the Davis manifold gluing
in the hyperplane $x_3=0$ and then in the equatorial plane. 
Thus $M$ is a hyperbolic gravitational instanton,   
with connected initial hypersurface $\Sigma$,   
and $M$ has a symmetry that maps $\Sigma$ onto 
a hypersurface $\Sigma'$ that is perpendicular to $\Sigma$. 
Thus $M$ is also a hyperbolic gravitational instanton
with connected initial hypersurface $\Sigma'$. 

The manifold $M$ can be constructed by gluing together two copies of 
the 120-cell $P$. 
Therefore the volume of $M$ is twice that of the Davis manifold and so  
the Euler characteristic of $M$ is 52.  
The homology groups of $M$ are    
$H_0(M) = \integers$, 
$H_1(M) = \integers_2^6\oplus\integers_4^2\oplus\integers^{18}$, 
$H_2(M) = \integers_2^6\oplus\integers_4^2\oplus\integers^{86}$, 
$H_3(M) = \integers^{18}$, and  
$H_4(M) = \integers$. 
The separating totally geodesic hypersurface $\Sigma$ of $M$ 
can be constructed by gluing together two copies of the fundamental 
domain for the Davis manifold cross-section in Figure~\ref{f-rat3}, 
and so the volume of $\Sigma$ is twice that of the cross-section 
of the Davis manifold.  
The volume of $\Sigma$ is approximately equal to $204.5$. 
The homology groups of $\Sigma$ are 
$H_0(\Sigma) = \integers$, 
$H_1(\Sigma)= \integers^{23}$, 
$H_2(\Sigma)= \integers^{23}$, and
$H_3(\Sigma)=\integers$.

\subsection{Noncompact Hyperbolic Gravitational Instantons}

In this section we relax the definition of a gravitational instanton 
by weakening the hypothesis of compactness to completeness with finite volume. 
Thus a gravitational instanton is now a complete, orientable, 
Riemannian 4-manifold $M$ of finite volume,  
satisfying Einstein's equations, 
with a separating, totally geodesic, orientable hypersurface $\Sigma$ 
which is the set of fixed points of an orientation reversing 
isometric involution of $M$. 
We will only consider hyperbolic noncompact gravitational instantons. 

A noncompact hyperbolic $n$-manifold $M$ of finite volume has  
a compact $n$-dimensional submanifold $M_0$ with boundary 
such that $M - M_0$  is a disjoint union of cusps and  
each boundary component of $M_0$ is a closed flat $(n-1)$-manifold. 
Each cusp $C$ is a Cartesian product $N \times (0,\infty)$ 
where $N$ is a closed flat $(n-1)$-manifold and $(0,\infty)$ 
is the open interval from $0$ to $\infty$. 
The metric on $C$ is $e^{-t}g + dt^2$ where $t$ 
is in $(0,\infty)$ and $g$ is the flat metric on $N$. 
In particular, the volume of the flat cross-section $N\times \{t\}$ of $C$ 
decreases exponentially as $t \to \infty$.  
This allows $C$ to have finite volume even though $C$ is unbounded.

In our paper [7], we constructed examples of noncompact hyperbolic
4-manifolds of smallest volume, that is, of Euler characteristic 1. 
Our examples were constructed by gluing together the sides of
a regular ideal 24-cell in hyperbolic 4-space. 
Our examples have totally geodesic hypersurfaces 
that are the set of fixed points of an isometric involution. 
This led G.W.Gibbons [3] to suggest that our examples 
may have applications in cosmology.  

A {\it regular {\rm 24}-cell} is a 4-dimensional, regular, convex, polytope with 24 sides, 
each a regular octahedron. Each side meets its eight neighbors 
along a triangular ridge.  Each edge of the 24-cell is shared by three sides, 
and each vertex is shared by six sides. 
There are a total of 96 ridges, 96 edges, and 24 vertices in a regular 24-cell. 
A {\it hyperbolic, ideal, regular {\rm 24}-cell} is a regular 24-cell 
in hyperbolic 4-space with all its vertices on the sphere at infinity 
(i.e. all vertices are ideal). 
The dihedral angle between adjacent sides of a regular ideal 24-cell is $\pi/2$. 

Let $Q$ be a hyperbolic, ideal, regular 24-cell. We realize $Q$ in the conformal 
ball model of hyperbolic 4-space with center at the origin and aligned 
so that the ideal vertices of $Q$ are 
$(\pm 1,0,0,0)$, $(0,\pm 1,0,0)$, $(0,0,\pm 1,0)$, 
$(0,0,0,\pm 1)$, and $(\pm 1/2,\pm 1/2,\pm 1/2,\pm 1/2)$. 
Then the four coordinate hyperplanes of $E^4$, 
given by $x_i = 0$, for $i = 1,2,3,4$, are planes of symmetry of $Q$. 
Let $K$ be the group of orthogonal transformations of $E^4$ 
generated by the reflections in the coordinate hyperplanes of $E^4$. 
Then $K$ is an abelian group of order 16 all of whose nonidentity 
elements are involutions. 

Our examples of noncompact hyperbolic 4-manifolds of Euler characteristic 1 
are obtained by gluing together the sides of $Q$ in such a way that 
each side $S$ of $Q$ is paired to a side $S'$ of $Q$ 
which is the image of $S$ under an element of $K$. 
The side-pairing map from $S$ to $S'$ is the composition of an element of $K$ 
that maps $S$ to $S'$ followed by the reflection in the side $S'$. 
In our paper [7], we computed that exactly 1171 nonisometric hyperbolic 4-manifolds 
can be constructed by such side-pairings of $Q$.   
All of these side-pairings of $Q$ are invariant under the group $K$. 
This implies that each coordinate hyperplane cross-section of $Q$ extends 
in each of our examples to a totally geodesic hypersurface  
which is the set of fixed points of an isometric involution. 
We call these hypersurfaces of our examples {\it cross-sections}. 

\begin{figure} 
\centerline{\epsfig{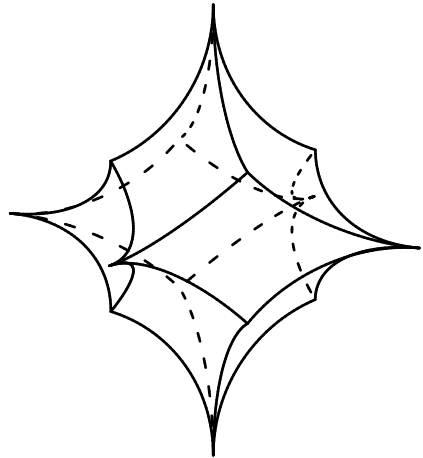}}
\caption{A hyperbolic right-angled rhombic dodecahedron}
\label{f-rat4}
\end{figure}

The intersection of a coordinate hyperplane of $E^4$ with $Q$ 
is a hyperbolic rhombic dodecahedron with dihedral angles $\pi/2$. 
A {\it rhombic dodecahedron} is a quasiregular convex polyhedron 
with 12 congruent rhombic sides. In a rhombic dodecahedron 
four rhombi meet at each vertex with acute angles and three rhombi 
meet at each vertex with obtuse angles. 
A hyperbolic rhombic dodecahedron with dihedral angles $\pi/2$ 
has ideal order 4 vertices. See Figure~\ref{f-rat4}. 

The cross-sections of our examples can be obtained 
by gluing together the sides of the rhombic dodecahedron in 
Figure~\ref{f-rat4}. 
In our paper [7], we classified all the possible cross-sections. 
It turns out that there are exactly 13 nonisometric cross-sections. 
In Table 1 we list all the data that we derived 
about these noncompact hyperbolic 3-manifolds. 

\begin{table*}
{\tentt
\tabskip 20pt plus 5pt
\halign to \linewidth{\tabskip 5pt plus 5pt
\hfil{\eighttt #}&\hfil #\hfil&\hfil #\hfil&\hfil #\hfil&\hfil #&
\hfil #\hfil&\hfil #\hfil&#\hfil \tabskip 20pt plus 5pt
&\tabskip 5pt plus 5pt
\hfil{\eighttt #}&\hfil #\hfil&\hfil #\hfil&\hfil #\hfil&\hfil #&
\hfil #\hfil&\hfil #\hfil&#\hfil\tabskip 20pt plus 5pt
\cr
$N$&$SP$&$O$&$C$&$S$\hfil&$H_{1}$&$H_{2}$&\hfil $LT$&
$N$&$SP$&$O$&$C$&$S$\hfil&$H_{1}$&$H_{2}$&\hfil $LT$\cr
\vrule height 16pt width 0pt 
1&142&1&3&48&300&2&TTT&
8&157&0&3& 8&201&1&KKT\cr
2&147&1&3&16&300&2&TTT&
9&367&0&3& 8&102&0&KKK\cr
3&143&0&3& 8&300&2&KTT&
10&174&1&4&64&400&3&TTTT\cr
4&156&0&3& 8&300&2&KTT&
11&134&0&4&16&310&2&KKTT\cr
5&357&0&3&16&220&1&KKT&
12&165&0&4& 8&220&1&KKKT\cr
6&136&0&3& 8&220&1&KKT&
13&135&0&4&16&121&0&KKKK\cr
7&153&0&3&16&201&1&KKT&
&&&&&&&\cr
}}
\caption{Cross-sections of the Ratcliffe-Tschantz hyperbolic %
4-manifolds}%
\end{table*}

The column of Table 1 headed by $N$ counts the manifolds.
The column headed by $SP$ describes the side-pairing
of the rhombic dodecahedron in a coded form that is explained in our paper [7]. 
We shall use the side-pairing code to identify a manifold in Table 1. 
The column headed by $O$ indicates the orientability of the manifolds
with 1 for orientable and 0 for nonorientable.
Note that only three of the manifolds are orientable, 
namely manifolds 142, 147, and 174. 
These three orientable manifolds are topologically equivalent 
to the complement of a link in the 3-sphere $S^3$. 
The manifolds 142, 147, and 174 are equivalent to the complement 
of the links $6_2^3$ (Borromean rings), $8_9^3$, and $8_2^4$, respectively.

The column of Table 1 headed by $C$ lists the number of cusps of the manifolds.
The link (flat cross-section) of each cusp is either a torus or a Klein bottle.
The column headed by $LT$ indicates the link type of each cusp
with T representing a torus and K a Klein bottle.
The column headed by $S$ lists the number of symmetries of the manifold.
The column headed by $H_1$ lists the first homology groups of the manifolds
with the 3 digit number $abc$ representing
$\integers^a \oplus \integers_2^b \oplus \integers_4^c$.
The column headed by $H_2$ lists the second homology groups of the manifolds
with the entry $a$ representing $\integers^a$.

The volume of the hyperbolic, right-angled, rhombic dodecahedron is 
$$8L(2) = 7.3277247\ldots,$$
where $L(s)$ is the Dirichlet $L$-function defined by 
$$L(s) = 1 - \frac{1}{3^s} + \frac{1}{5^s} - \frac{1}{7^s} + \cdots.$$
All the manifolds in Table 1 have the same volume as the  
right-angled rhombic dodecahedron, since they are constructed by 
gluing together the sides of the rhombic dodecahedron.

Only 22 of the 1171 hyperbolic 4-manifolds constructed in our paper [7] 
are orientable. Table 2 lists all the data that we derived for 
these 22 noncompact, orientable, hyperbolic 4-manifolds. 

\begin{table*}
{\tentt 
\tabskip 20pt plus 5pt
\halign to\linewidth{\tabskip 5pt plus 5pt
\hfil{\eighttt #}&\hfil #\hfil&\hfil #\hfil&\hfil #\hfil&\hfil #\hfil&
\hfil #\hfil&\hfil #\hfil\tabskip 10pt plus 5pt&
\tabskip 5pt plus 5pt\hfil #\hfil&\hfil #\hfil&
\hfil #\hfil&\hfil #\hfil\tabskip 20pt plus 5pt
\cr
$N$&$SP$&$S$&$H_{1}$&$H_{2}$&$H_{3}$&$LT$&
$CS_{1}$&$CS_{2}$&$CS_{3}$&$CS_{4}$\cr
\vrule height 16pt width 0pt 
1&1428BD&16&330&700&4&AAABF&156-1&174-2&147-2&142-2\cr
2&14278D&16&240&600&4&AABBF&146-1&173-1&134-1&142-2\cr 
3&1477B8&16&240&600&4&AABBF&156-1&143-1&137-1&147-2\cr
4&1477BE&16&240&600&4&AABBF&357-1&153-1&137-1&147-2\cr
5&1478ED&16&240&600&4&AABBF&357-1&174-2&146-1&147-2\cr
6&14278E&16&240&600&4&ABBBF&147-2&153-1&134-1&142-2\cr
7&142DBE&48&150&500&4&ABBBF&157-1&157-1&157-1&142-2\cr
8&1427BD&16&150&500&4&ABBBF&156-1&173-1&137-1&142-2\cr
9&1477EB&16&150&500&4&ABBBF&367-1&163-1&136-1&147-2\cr
10&1477ED&16&150&500&4&ABBBF&357-1&173-1&136-1&147-2\cr
11&1478EB&16&150&500&4&ABBBF&367-1&134-1&146-1&147-2\cr
12&147BDE&16&150&500&4&ABBBF&367-1&156-1&175-1&147-2\cr
13&14B8ED&16&150&500&4&ABBBF&367-1&174-2&146-1&143-1\cr
14&1427BE&16&150&500&4&BBBBF&157-1&153-1&137-1&142-2\cr
15&1477DE&16&150&500&4&BBBBF&367-1&153-1&135-1&147-2\cr
16&14B7E8&16&060&400&4&BBBBF&175-1&143-1&136-1&143-1\cr
17&14B7ED&16&060&400&4&BBBBF&367-1&173-1&136-1&143-1\cr
18&14BDE7&16&060&400&4&BBBBF&567-1&137-1&156-1&143-1\cr
19&14B7DE&16&060&400&4&BBFFF&567-1&153-1&135-1&143-1\cr
20&14B8E7&16&051&400&4&ABFFF&567-1&134-1&146-1&143-1\cr
21&14BD7E&16&051&400&4&ABFFF&537-1&157-1&153-1&143-1\cr
22&17BE8D&16&051&400&4&ABFFF&153-1&367-1&134-1&173-1\cr
}}
\caption{The orientable, Ratcliffe-Tschantz, hyperbolic 4-manifolds}%
\end{table*}

The column headings in Table 2 are as in Table 1. 
All 22 manifolds in Table 2 have five cusps. 
The column headed by $LT$ lists the link types of the cusps, 
where ${\rm A} = O_1$ is the 3-torus, ${\rm B} = O_2$ is the half-twisted 3-torus, 
and ${\rm F} = O_6$ is the Hantzsche-Wendt 3-manifold [5].
The column headed by $CS_i$ gives the cross-section of the manifolds 
determined by the coordinate hyperplane $x_i=0$. 
Here the -1 refers to a one-sided cross-section and -2 refers to 
a two-sided cross-section.  
It is worth noting that a hypersurface of an orientable manifold 
is two-sided if and only if the hypersurface is orientable. 

Let $N$ be an orientable hyperbolic 4-manifolds in Table 2 
and let $S$ be a one-sided cross-section of $N$. 
Then $S$ is nonorientable.   
Let $M_R$ be the manifold with boundary obtained by cutting $N$ along $S$. 
Then $M_R$ is a connected, orientable, hyperbolic 4-manifold 
with a totally geodesic boundary $\Sigma$ equal to the orientable double cover of $S$. 
Let $M$ be the double of $M_R$. Then $M$ is a noncompact, hyperbolic,  
gravitational instanton with connected initial hypersurface $\Sigma$. 
The volume of $\Sigma$ is twice the volume of $S$, and so 
the volume of $\Sigma$ is 
$$16L(2) = 14.6554494\ldots.$$
The manifold $M$ is a double cover of $N$; therefore 
the Euler characteristic of $M$ is twice that of $N$, 
and so $\chi(M) = 2$.     
Thus every manifold in Table 2 has a double cover which 
is a noncompact, hyperbolic, gravitational instanton of smallest possible volume.  

Let $N$ be one of the 1149 nonorientable hyperbolic 4-manifolds 
constructed in our paper [7] and let $S$ be a cross-section of $N$. 
Then $S$ does not separate $N$, since the Euler characteristic of $N$ is odd. 
Let $M$ be the orientable double cover of $N$ and 
let $\Sigma$ be the hypersurface of $M$ covering $S$. 
Then $M$ is a gravitational instanton with initial surface $\Sigma$ 
if and only if $\Sigma$ is connected and separates $M$, 
since the reflective symmetry of $N$ along $S$ lifts 
to a reflective symmetry of $M$ along $\Sigma$. 

Suppose that $\Sigma$ is connected and separates $M$. 
Then $\Sigma$ is two-sided in $M$. 
Therefore $\Sigma$ is orientable, since $M$ is orientable. 
Let $V$ be a regular neighborhood of $S$ in $N$  
which is invariant under the reflective symmetry of $N$ along $S$. 
Then $V$ lifts to a regular neighborhood $U$ of $\Sigma$ in $M$  
which is invariant under the reflective symmetry of $M$ along $\Sigma$.  
Now $U$ is the Cartesian product of an open interval and $\Sigma$. 
The complement of $U$ in $M$ is the union of two disjoint 
connected manifolds $M_1$ and $M_2$ with boundary homeomorphic to $\Sigma$. 
Let $N_0 = N - V$. Then $N_0$ is a connected manifold, since $S$ 
does not separate $M$, and the boundary of $N_0$ is the boundary of $V$. 
The manifolds $M_1$ and $M_2$ are homeomorphic to $N_0$, 
since  $M-U$ double covers $N-V$. 
Therefore the boundary of $V$ is homeomorphic to $\Sigma$. 
Hence $S$ is one-sided, since the boundary of $V$ is connected. 
Now $S$ must be orientable since otherwise $V$ would be a twisted $I$-bundle, 
and hence orientable, but then $V$ would be evenly covered, and so 
$\Sigma$ would be disconnected which is not the case. 
Thus $S$ must be orientable and one-sided. 

Conversely, if $S$ is orientable and one-sided, 
then $\Sigma$ is connected and two-sided in $M$, 
since a regular neighborhood of $S$ in $N$ is nonorientable. 
Moreover, $\Sigma$ separates $M$ if and only if the complement of 
$S$ in $N$ is orientable, since $M-\Sigma$ double covers $N-S$. 
Thus the orientable double cover $M$ of $N$ is a gravitational instanton,  
with connected initial hypersurface $\Sigma$ covering the cross-section $S$ of $N$,  
if and only if $S$ is orientable, one-sided, 
and the complement of $S$ in $N$ is orientable.

We now describe an explicit example of a noncompact, hyperbolic, 
gravitational instanton $M$ obtained as the orientable 
double cover of one of the nonorientable hyperbolic 4-manifolds $N$ 
of Euler characteristic 1 constructed in our paper [7]. 
Let $e_1,e_2,e_3,e_4$ be the standard basis vectors of $E^4$. 
Then the 24 ideal vertices of the 24-cell $Q$ are $\pm e_1,\pm e_2,\pm e_3,\pm e_4$, 
and $\pm\frac{1}{2}e_1\pm\frac{1}{2}e_2\pm\frac{1}{2}e_3\pm\frac{1}{2}e_4$. 
The 24 sides of $Q$ are regular ideal octahedra 
lying on unit 3-spheres in $E^4$ centered at the points $\pm e_i\pm e_j$. 
A pair of distinct vertices $\{\pm e_i,\pm e_j\}$ 
from $\{\pm e_1,\pm e_2,\pm e_3,\pm e_4\}$, 
which are not antipodal, determines a unique side of the 24-cell 
having this pair as vertices and it will be
convenient to refer to this side by the center 
$\pm e_i\pm e_j$ of the 3-sphere containing this side.
The group of orthogonal transformations of $E^4$ generated by the 
reflections in the coordinate hyperplanes of $E^4$ can be identified 
with the group of orthogonal $4\times 4$ diagonal matrices,  
$$K=\{{\rm diag}(\pm 1,\pm 1, \pm 1, \pm1)\}.$$
We describe the manifold $N$ by specifying a gluing of the
24-cell $Q$, the gluing defined by side-pairing maps of $Q$.
A side-pairing map will be specified by an element of $K$ mapping
a side $S$ to another side $S'$ followed by reflection in $S'$.
The ridges will have to be matched in cycles of 4 and the edges in cycles
of 8 in order to define a hyperbolic 4-manifold.

Take $e_4$ as the north pole, $-e_4$ as the south pole, and the 
coordinate hyperplane $x_4=0$ as the equatorial plane of our 24-cell $Q$.  
Take side-pairing maps induced by elements of $K$ as follows.  For sides
centered at $\pm e_1\pm e_2$ take ${\rm diag}(1,-1,1,-1)$, for sides centered
at $\pm e_2\pm e_3$ take ${\rm diag}(1,1,-1,-1)$, and for sides centered
at $\pm e_3\pm e_1$ take ${\rm diag}(-1,1,1,-1)$, permuting cyclically in the
first three coordinates to define the side-pairings of the sides
perpendicular to the equatorial plane.  For sides centered at
$\pm e_1\pm e_4$ take ${\rm diag}(1,1,-1,-1)$, for sides centered at
$\pm e_2\pm e_4$ take ${\rm diag}(-1,1,1,-1)$, and for sides centered at
$\pm e_3\pm e_4$ take ${\rm diag}(1,-1,1,-1)$, preserving the cyclic symmetry
in the first three components to define the side-pairings of the sides
not intersecting the equatorial plane other than at an ideal vertex.
Then we can check that the ridges are in cycles of 4 and the edges
are in cycles of 8 and so we get a hyperbolic 4-manifold $N$ 
(isometric to the manifold 1096, with side-pairing code 56CC65, in our paper [7]).  
Because the last coordinate is flipped by each of the symmetries, 
sides in the northern half of $Q$ are paired with sides in the southern half, 
and northern halves of sides perpendicular to the equatorial plane 
are paired to southern halves of sides.  
Each side-pairing map is an orientation preserving (determinate
$+1$) symmetry of $Q$ followed by reflection in a side and as such
is orientation reversing.  Restricted to the equatorial plane however,
the side-pairing maps of the right-angled rhombic dodecahedron
are orientation preserving.  Thus the equatorial cross-section in $N$ 
is an orientable totally geodesic hypersurface $S$ which is one-sided in $N$.  
The cross-section $S$ is isometric to the manifold 142 (Borromean rings complement) 
in Table 1. 

The orientable double cover $M$ of $N$ can be
described then by a corresponding gluing of two copies of the 24-cell $Q$,
taking the same pairings of sides but crossing between the two copies.
Thus the northern half of one 24-cell is always glued to the southern
half of the other 24-cell.  The equatorial cross-sections of the two
24-cells thus glue up to a double cover $\Sigma$  of $S$ 
which is a separating, totally geodesic, hypersurface  
which is also a mirror for the orientable 4-manifold $M$. 
Thus $M$ is a noncompact hyperbolic gravitational instanton 
with connected initial hypersurface $\Sigma$. 

The Euler characteristic of $M$ is twice that of $N$, and so $\chi(M) = 2$. 
Thus $M$ is a noncompact hyperbolic gravitational instanton 
of smallest possible volume. 
The manifold $M$ has $H_1(M)=\integers_2^3\oplus\integers_4^2\oplus\integers^3$,
$H_2(M)=\integers^{12}$ and $H_3(M)=\integers^8$.
Its equatorial cross-section $\Sigma$ has
$H_1(\Sigma)=\integers_2^2\oplus\integers^3$ and $H_2(\Sigma)=\integers^2$.
The nonorientable hyperbolic 4-manifold $N$ has 6 cusps, 
3 along the equatorial plane corresponding to the three cusps of $S$, 
and 3 off the equatorial plane.  
The orientable double cover $M$ has 9 cusps, the cross-section $\Sigma$ still
has 3 cusps, but the original 3 cusps off of the equatorial plane are double 
covered to give 3 cusps on each side of $\Sigma$. 
The volume of $\Sigma$ is twice the volume of $S$, and so 
the volume of $\Sigma$ is 
$$16L(2) = 14.6554494\ldots.$$
The manifold $M$ is but one of many examples of hyperbolic, noncompact, gravitational  
instantons of smallest possible volume that arise as the orientable double cover 
of one of the 1149 nonorientable hyperbolic 4-manifolds of Euler characteristic 1 
constructed in our paper [7]. 

\newpage
\centerline{References}

\begin{enumerate}
\item Coxeter, H. S. M., {\it Regular Polytopes}, Third Edition, 
Dover, New York, 1973.

\item Davis, M. W., A hyperbolic 4-manifold,
{\it Proc. Amer. Math. Soc.}, 93 (1985), 325-328.

\item Gibbons, G. W., Tunnelling with a negative cosmological constant, 
{\it Nuclear Physics B}, 472 (1996), 683-708.

\item Gromov, M. and Piatetski-Shapiro, I., 
Non-arithmetic groups in Lobachevsky spaces,
{\it Inst. Hautes \'Etudes Sci. Publ. Math.}, 66 (1988), 93-103.

\item Hantzsche, W. and Wendt, H., Dreidimensionale euklidische Raumformen,
{\it Math. Ann.}, 110 (1935), 593-611.

\item Ratcliffe, J. {\it Foundations of Hyperbolic Manifolds},
Graduate Texts in Math., vol. 149, Springer-Verlag, Berlin, Heidelberg, and
New York, 1994.

\item Ratcliffe, J. and Tschantz, S., The volume spectrum of hyperbolic 4-manifolds, 
Experimental Math. 9 (2000), 101-125.

\item Wolf, J. A., {\it Spaces of Constant Curvature}, 
Fifth Edition, Publish or Perish, Wilmington, DE, 1984.

\end{enumerate}




\renewcommand\Box{\protect\mbox{\,\protect\fbox{\protect\rule{0ex}{0.9ex}\protect\rule{0.9ex}{0ex}}\,}}

\def\fp {fundamental polyhedron}
\def\uc {universal covering}
\def\flz {Fang \cite{Fan90}}
\def\sos {Sokoloff \& Shvartsman \cite{Sok74}}
\def\fw {Fagundes \& Wichoski \cite{Fag87}}
\def\lelalu {Lehoucq \etal (1994)}
\def\ks {Kantowski--Sachs}

\subtitle{Topology, the vacuum and the cosmological constant}

\subauthor{Marc Lachi\`eze-Rey}
\subaddress{CNRS URA - 2052
\\CEA, DSM/DAPNIA/ Service d'Astrophysique
\\CE Saclay, F-91191 Gif--sur--Yvette CEDEX, France}
\date{\ }

%

\def\Ren {\Re ^n}
\def\lr{Lachi\`eze-Rey}
\def\ccc {cosmological constant}
\def\npage {\vfill \eject}
\def\mc {multi-connected}
\def\bbg {big bang}
\def\cct {cosmological constant}
\def\Re{{\rm I\!R}}
\def\bbbr{{\rm I\!R}} 
\def\bbbrn{{\rm I\!R}^{n}} 
\def\bbbm{{\rm I\!M}}
\def\bbbn{{\rm I\!N}} 
\def\bbbf{{\rm I\!F}}
\def\bbbs{{\rm S}}
\def\bbbh{{\rm I\!H}}
\def\bbbk{{\rm I\!K}}
\def\bbbp{{\rm I\!P}}
\def\bbbz{{\rm \!Z}}
 \def\etc {{\it etc.}}
 \def\apriori {{\it a priori}}
\def\fwhm  {{\sc fwhm}}
\def\kms {\mbox{km~s}^{-1}}
\def\hmpc {h^{-1}~\mbox{Mpc}}
\def\kmsmpc {\mbox{km ~ s}^{-1}~\mbox{Mpc}^{-1}}
\def\mpc {\mbox{Mpc}}
\def\exp {\mbox{exp}}
\def\cos {\mbox{cos}}
\def\sin {\mbox{sin}}
\def\cosh {\mbox{cosh}}
\def\sinh {\mbox{sinh}}
\def\squareforqed{\hbox{\rlap{$\sqcap$}$\sqcup$}}
\def\sq{\ifmmode\squareforqed\else{\unskip\nobreak\hfil
\penalty50\hskip1em\null\nobreak\hfil\squareforqed
\parfillskip=0pt\finalhyphendemerits=0\endgraf}\fi}
 \def\uc  {universal covering}
\def\etal {et al.~}
\def\bm {\boldmath}
\def\deg {^\circ}
\def\sqdeg {^{\circ2}}
\def\sol {_{\odot}}
\def\msol {M_{\odot}}
 \def\dS{de~Sitter}
\def\eds{Einstein -- de~Sitter}
\def\frl{Friedmann-Lema\^ \i tre}
\def\lemaitre {Lema\^ \i tre}
\def\sz {Sunyaev-Zeldovich}
\def\kg {Klein-Gordon}
\def\sw{Sachs-Wolfe}
\def\rms {{\it rms}}
\def\mink {Minkowski spacetime}
\def\elm {electromagnetic}
\def\eg {{e.g.}}
\def\ie {{i.e.}}
\def\pp {{\bf p}}
\def\PP {{\bf P}}
\def\lss {last scattering surface}
\def\gr {general relativity}
\def\rec {recombination}
\def\lg {Local Group}
\def\mw {Milky Way}
\def\cmb {Cosmic Microwave Background}
\def\pd {probability distribution}
\def\df {distribution function}
\def\spt {space-time}



\submaketitle

\begin{abstract}
If  the topology of
space is multi-connected, rather than simply connected as it is most
often assumed, this would cause  a major revolution in
cosmology, and  a huge progress in the knowledge of our world (see the
review paper by   Lachi\`eze-Rey \& Luminet, 1995, hereafter LaLu).
 This  would  set  new constraints and ask new questions  on the physics of
the
primordial universe.  Why space is
multi-connected~? What has determined its  principal
directions    and the values of its spatial dimensions~?
The links between  cosmology and  quantum physics would be modified,
in particular    the question  of the vacuum energy and
of the cosmological constant.
\end{abstract}

\subsection {Introduction}

Many aspects of topology   concern cosmology and theoretical physics.
For instance, some work in  quantum gravity or in the search for
fundamental interactions (see, for instance, Spaans 1999 and  Rovelli 1999)
suggest
  that the topology of spacetime at the
microscopic scale may be different than that of $\Ren$. At the
macroscopic scale, speculative ideas in quantum cosmology
(Ellis, 1975; Atkatz \& Pagels, 1982;
Zel'dovich \& Starobinsky, 1984; Goncharov \& Bytsenko, 1989) seem to favor
the \mc ~case. Topological transitions, forbidden in classical \gr,  are
allowed
in quantum cosmology. A "~spontaneous birth" of the universe is
sometimes claimed to lead  "~probably~"   to a  \mc ~universe.

  Some  theories (Klein, 1926, 1927;  Thiry, 1947;  Souriau, 1963, \ldots,
up to
superstrings), introduce
additional dimensions which are compactified, \ie, which  have a \mc ~topology.
If this is the  case, it would
appear rather natural that the dimensions of physical space are also \mc, even
if with a much larger scale. Here I consider only the possibility that  the
topology of
our three dimensional space is \mc ~(I  consider the
natural  topology linked to the spatial part of the    metric). This
implies that at least
one dimension of
space is closed, and in many cases, that space is of finite volume and
circumference. I   refer to an universe with multiconnected space as
a  {\sl small universe}.

\subsubsection{Topology and cosmology}

Observations are necessary  to decipher the topology of our
space. The case  is especially interesting today, given    the favorite value
 of $\Omega$, lower than 1,   which suggest a negative spatial
 curvature:   multiconnectedness would become  the only possibility for a
 closed (finite) space. For a review of the possible observational
 tests, see Lalu, and \lr ~1999. I assume  the  {\sl global
 hyperbolicity}  of \spt, implying     the
manifold structure of     $\cal M$$_3 \times \Re _{time} $.
I also impose     spatial orientability.  For a presentation of the main
geometrical
tools to handle   topology, see Lalu, or the reference books by
  Thurston (1978) and  Nakahara (1990).


\subsubsection{Characteristic lengths}

In any cosmic model with non zero spatial curvature,
the     curvature radius of space,
$R_{curv}=(c/H_{0})~/~\sqrt{1-\Omega -\lambda}$,
provides a natural length unit. It  determines the
possible sizes and shapes of a small universe.
On the other hand,
the   {\sl observable} universe is characterized by the Hubble length,
and    the  horizon radius $R_{horizon} \approx 2 R_{curv}~ Arctanh
\sqrt{1-\Omega -\lambda}$, with  the corresponding volume
$V_{horizon}=\frac{4\pi~R_{horizon}^{3}}{3}$.

A  relevant parameter to measure the degree of
visibility and relevance of the property of \mc ness is given
by $ B=  V_{horizon}/V$, where  $V$ is the spatial volume of the
small universe. I  call $r_{-}$ the   {\sl internal radius}, the radius of
the largest
        (geodesic) sphere in the fundamental   polyhedron, and   $r_{+}$
the {\sl external radius}, the radius of the smallest
sphere        in which the fundamental   polyhedron is inscribed.
A multiconnected space  with zero curvature may have arbitrary
dimensions. Those of a space with negative curvature are constrained
by the value of the (constant) curvature.

The   smallest space with {\sl negative}    curvature known today is the
{\sl Weeks space}, with volume
	$V= 0.9427$. Its \fp ~has    18 faces, with the values $r_{+} =
	0.7525$ and $r_{-} = 0.5192$. The {\sl Thurston space} (Thurston
1982) has
$V=0.9814$.	 The {\sl  cylindrical horn space}, studied by Sokolov and
Starobinsky, is non compact.

\subsection{Topology and vacuum energy}

The multi-connectedness of space  modifies
the limiting conditions of the universe, more precisely here, of space.
They modify   the calculations of the  classical  or
quantum fields,   in particular of  their fundamental
state, or " vacuum ", and of its stress-energy tensor.
A conseqence is the    possibility  of  some "~topological Casimir
effect~" (Mostepanenko and Trunov, 1988).

This is based on the
(still speculative) idea that   " vacuum energy " and pressure
may exert some  gravitational effects at the cosmic scale. Those are
for instance  often invoked to give rise to an    inflationary era,  or
to some  peculiar cosmic dynamics. Very often, they are claimed  to mimic a
cosmological constant.

A true  cosmological constant  $\Lambda = 3~H_{0}^{2}~\lambda$
(different from a  vacuum energy) may be present. This  is allowed in
(some versions of) \gr, but there is
no natural scale for it. Although its non zero value  would  remain
unexplained, there is no "~\ccc ~problem~": the expression refer in fact to
a "~vacuum energy problem~", since there is a natural scale for
vacuum energy (coming from particle physics) in contradiction with
cosmological observations.
A cosmic length $L_{\Lambda} = \frac{ 3000
~\hmpc}{\sqrt{\lambda }}$ is associated to  $\Lambda$. Since it may be of
the same
magnitude order than  the
lengths associated to a small universe, this motivates
examination of possible effects which could mimic such a constant in
a small universe. Let us emphasize  that   vacuum energy and  cosmological
constant are conceptually different, and  also have different consequences
onto
the cosmic evolution, excepted in the case of  \mink.

In Minkowski \spt,    quantum field theory  associates a momentum energy
tensor
$T_{\mu \nu } = - \rho _{V} ~ g_{\mu \nu }$    to the fundamental state of
a (scalar)
field. Its  gravitational interaction and   cosmological
effects, if any, would be  analog to that of   a perfect fluid with density
$  \rho_{V}$,
and pressure $P= -\rho_{V}$. This corresponds to an index
$\gamma=0$, and a dilution law  $\rho_{V} \propto Cte$ in time. This  is
also  analog to the  effect  of  a cosmological constant.
Similar effects are also expected for a  universe whose  dynamics is
dominated by   a {\sl scalar field}  $\phi$ (again, in Minkowski
\spt), with  momentum - energy tensor
\begin{equation}
T_{\mu \nu } =\phi _{,\mu} ~\phi _{,\nu} - \eta _{\mu \nu} [1/2 \phi
_{,\rho}~ \phi ^{,\rho} + V(\phi) ].
\end{equation}

For a  field   constant in space and  time ($ \phi _{,\mu} =0$), this
reduces to
$T_{\mu \nu }=V(\phi)~ g_{\mu \nu }$. Also, if  $\phi
_{,\rho}~ \phi ^{,\rho} <0$, this is analog to a perfect
fluid with density  $\mu = 1/2 ~\dot{\phi }^{2} +V,$ and pressure
$P = 1/2 ~\dot{\phi }^{2} -V$.  These formulae should be
extended  to curved, expanding and,   here,  \mc ~\spt.

\subsubsection{Quantum fields in non Minkowskian \spt}

A a scalar field  obeys the classical equation  ,
\begin{equation} \label{classical}
 {\cal O} \phi \equiv  (\Box+m^{2}) \phi =0,
\end{equation}  deriving  from the   Lagrangian $
{\cal L}  = 1/2~(\phi _{,\mu} ~\phi ^{,\mu}-m^{2}~ \phi^{2})$.
Usual  quantification (in \mink) proceeds through the following steps:
\begin{itemize}
\item select a set of positive frequency  orthogonal modes $u_{k}$,
solutions of the classical equation  (\ref{classical}),
\item quantize the modes, by introducing the conjugated  moments
 $\Pi \equiv \frac{\partial {\cal L}}{\partial(\partial
_{t \phi)}}$, which obey  the commutation relations at equal times
:\\$[\phi,\phi]=[\Pi,\Pi]=0$, and $[\phi,\Pi]=\delta;$
\item decompose any field  over the modes  \begin{equation}
\phi  = \sum a_{k }~u_{k} + a_{k }^{+} ~u_{k}^{*}.
\end{equation}
\item This gives the equivalent commutation relations : \\
  $[a_{k },a_{k' }]=[a_{k }^{+},a_{k' }^{+}]=0$
  \item The creation and annihilation operators   define the vacuum
  state  $\mid 0 > $, such that   $[a_{k } \mid 0 > =0$.
\item Its impulsion and energy are given by \\
$<0 \mid \PP \mid 0 >$ and $<0 \mid H \mid 0 >$, with
$H=\int T_{tt}~ dV$ and $P_{i}=\int T_{ii} ~dV$.
\end{itemize}

Extension of this procedure (originally  defined in \mink) to curved, or \mc
~\spt ~is considered, for instance, in Birrel and Davies (1982):   \spt
~curvature (spatial curvature and expansion)
modifies  the modes.  Multi-connectedness modifies the  limiting
conditions and
restricts the admissible   modes.

For   a classical field with  the field equations
$(\Box+m^{2} + \xi R) \phi =0$ ($\Box$   is the d'Alembertian in
 curved \spt,
$R$ the   Ricci scalar, and  $\xi$ a (conformal)  coupling), the proper
modes are in general  non  covariant, and  depend on the   coordinates
system. The vacuum, obtained from the quantization procedure
depends on the choice of the proper modes. Applied to fields in the
vicinity of
black holes, this gives their associated
temperature  $T= \frac{1}{8\pi~k~M}$; in de Sitter space, this gives  a
temperature
 $T= \frac{1}{8\pi~k~a}$, $a$
being  the \spt  ~curvature radius. An
accelerated observer  (Rindler \spt), looking at the inertial vacuum,
sees   a temperature  $T= \frac{a}{2\pi~k}$ (the Unruh effect).

\subsubsection{Topological Casimir effect}

Birrel and Davis  (1982) calculate, as an illustration,    the vacuum
for a
two dimensional static  cylindrical  universe, with  circumference   $L$.
For any field, and thus for the modes, the cylindricity condition
reads
$U_{k}(x) = U_{k}(x+L)$ (periodical), or $U_{k}(x) = U_{k}(x+nL)$ (twisted).
This restricts   the    possible modes and   modifies in consequence   the
vacuum and
the associated momentum-energy tensor:  instead of modes
$U_{k}(x) = \frac{1}{2\omega} e^{ikx-i\omega t}$, with $k$
arbitrary, they lead to the modes   $U_{k}(x) = \frac{1}{2L\omega} e^{i2\pi
nx/L-i\omega t}$, with $n$ an integer.
The result is a perfect fluid contribution, with density and
pressure $\rho = + p = \frac{-\pi}{L^{2}}$.  The
density appears to    scale  $\propto \frac{1}{L^{2}}$, like that
associated to a cosmological
constant.  The stress-energy tensor, however,    does not identify with a
cosmological constant term.

We have generalized this calculation in a   3+1 dimensional
\spt ~ $\Re ^{2} \times S _{z}  \times
\Re_{time}$, with  adiabatic approximation (static space), to a scalar field,
with zero mass and no coupling.
The result is a density $\rho = <T_{00}>= \frac{-\pi ^{2}}{90 L^{4}}$, and
other components of the momentum energy tensor as
$$ <T_{xx}>=<T_{yy}>= <T_{00}>$$
$$ <T_{zz}>=3/2~<T_{00}>.$$

The tensor is not isotropic ($z$ is the closed dimension). We 
lose
the   analogy with   a perfect fluid   or  a   cosmological
constant term. We  also lose the
 $\frac{1}{L^{2}}$ scaling of the density.
Moreover, the numerical value obtained,
  $\rho =\frac{10^{-82}~g.cm^{-3}}{L^{4}_{Gpc}}$, is much  smaller
  than any value of   cosmological interest. This is, again, the
  vaccum energy problem, arising when one
tries to interpret  the cosmological constant as a  particle physics
(here a quantum field)
effect.

By analogy, corresponding
calculations have been made  for an     hypertorus, with
result also  different from a cosmological constant.
Extensions to  \elm ~and fermionic fields are expected
to lead to smilar forms and
orders of magnitude.  In the non static case, the cosmic  expansion
makes the results  more complex, with a time evolution of the vacuum.
We obtained for   instance
$$t = \sqrt{\frac{ 360 \pi}{G}} ~\ln a +\frac{G}{14 a^{4}}~(\frac{ 360
\pi}{G})^{3/2},
~~a>>0.$$

Elizalde  and Kirsten (1994) and
 Goncharov (1982) have    calculated the cases of a toroidal space-time
with an
 arbitrary number of dimensions.  Bytsenko  and Goncharov (1991)
have obtained some partial results  for the  case with negative spatial
curvature.

\subsection{Conclusion}

The \mc ness of our universe remains a fascinating possibility,
favored by modern ideas in theoretical physics. Present observations
apparently exclude a \mc ~space much smaller than horizon, for
positive or null curvature. But space can be \mc, with a scale much
smaller than the horizon, if the space curvature is negative (a result
favored by recent observations).

Multiconnectedness (even   with a scale comparable to that of the
horizon, although this would be very difficult to recognize) would  lead
to very
interesting effects concerning  the development of the  fluctuations leading
to the formation of the large scale structures, and to the
anisotropies of the CMB; and also  the quantization of fields, with a
possible feedback onto the
dynamics of the universe.

Both kinds of effects thus deserve to be explored. In addition, it is
necessary to continue the efforts to detect a possible \mc ness of
space, especially in the case of negative spatial curvature.

\npage 




\newtheorem{acknowledgement}[theorem]{Acknowledgement}
\newtheorem{algorithm}[theorem]{Algorithm}
\newtheorem{axiom}[theorem]{Axiom}
\newtheorem{case}[theorem]{Case}
\newtheorem{claim}[theorem]{Claim}
\newtheorem{conclusion}[theorem]{Conclusion}
\newtheorem{condition}[theorem]{Condition}
\newtheorem{conjecture}[theorem]{Conjecture}
\newtheorem{criterion}[theorem]{Criterion}
\newtheorem{definition}[theorem]{Definition}
\newtheorem{example}[theorem]{Example}
\newtheorem{exercise}[theorem]{Exercise}
\newtheorem{notation}[theorem]{Notation}
\newtheorem{problem}[theorem]{Problem}
\newtheorem{proposition}[theorem]{Proposition}
\newtheorem{remark}[theorem]{Remark}
\newtheorem{solution}[theorem]{Solution}
\newtheorem{summary}[theorem]{Summary}
\newenvironment{proof}[1][Proof]{\textbf{#1.} }{\ \rule{0.5em}{0.5em}}
\newdimen\dummy
\dummy=\oddsidemargin
\addtolength{\dummy}{72pt}


\subtitle{Creation of a Closed Hyperbolic Universe}

\subauthor{S. S. e Costa and H. V.Fagundes}
\subaddress{Instituto de F\'{i}sica Te\'{o}rica,
Universidade Estadual Paulista\\
S\~{a}o Paulo, SP 01405-900, Brazil\\
\textit{e-mail:}
helio@ift.unesp.br\\
{\epsfxsize=11cm \epsfbox[72 310 540 470]{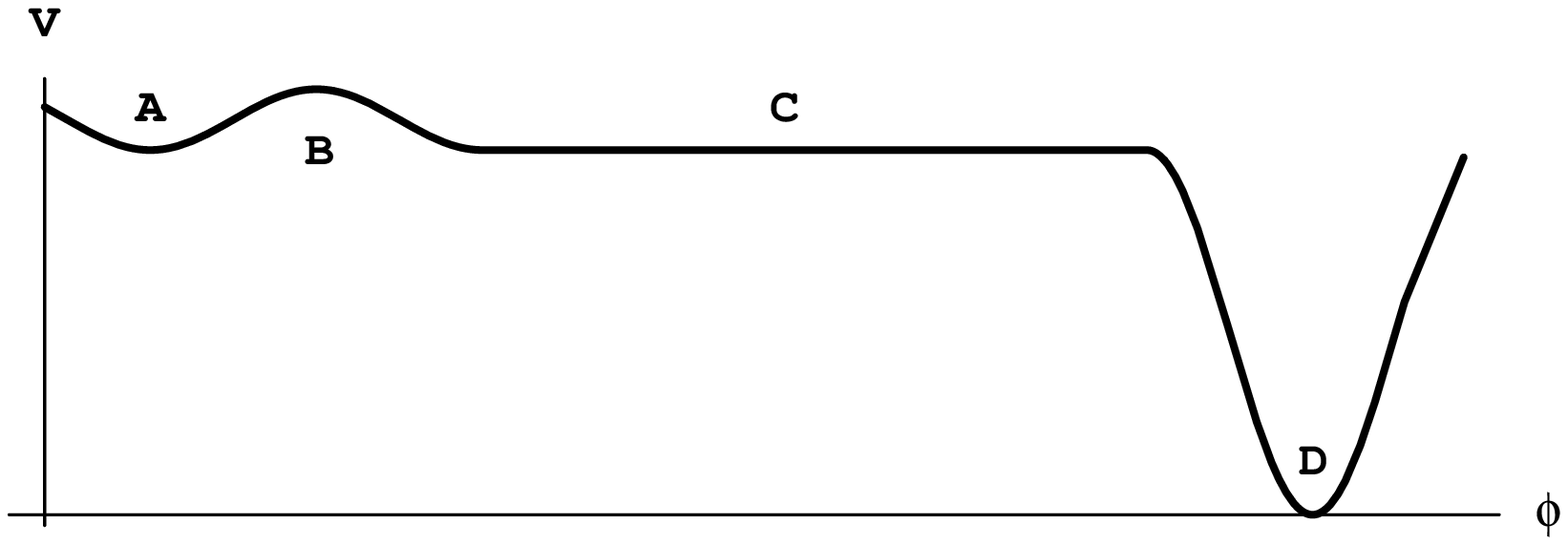}} \\
\addtocounter{figure}{1} 
Figure~\protect\arabic{figure}. Potential $V(\phi)$. 
\addcontentsline{lof}{figure}{\numberline
{\arabic{figure}}{\ignorespaces Potential 
$V(\phi)$}} 
}

\submaketitle



This short report is essentially based on our more extended paper \cite{FeC}.

We assume a primordial a real scalar field $\phi =\phi (t)$ and a potential $%
V(\phi )$ as in the figure above,
with a false vacuum at $\phi _{0}$ in region $A$ of
the figure. $V(\phi _{0})$ acts as a posititive cosmological constant; then
Wheeler-DeWitt's equation for a spherical, homogeneous and isotropic
universe leads to the spontaneous creation of a spacetime (cf. Gibbons \cite
{GWG}) $\mathcal{M=M}_{R}\cup \mathcal{M}_{L}$, where $\mathcal{M}_{R}$ is
one-half of de Sitter's instanton with topology $S^{4}$ and $\mathcal{M}_{L}$
is de Sitter's spherical spacetime with topology $\mathbf{R}\times S^{3}.$
The latter's scale factor is $R_{0}\cosh (t/R_{0})$

We now extend this process to topologies $S^{4}/\Gamma ,$ $\mathbf{R}\times
(S^{3}/\Gamma )$, respectively, where $\Gamma $ is a subgroup of the group
of isometries $Isom(S^{3})$ such that $S^{3}/\Gamma $ is a 3-spherical
manifold (see, for example, Lachi\`{e}ze-Rey and Luminet's review \cite{LaLu}%
), and $S^{4}/\Gamma $ is a 4-spherical orbifold \cite{Scott}. The idea is
to have a control over the volume (normalized to unity curvature) of the
created universe, which is $2\pi ^{2}/($order of $\Gamma ).$

Then we postulate a metric and topology change by a quantum process,
related to the potential barrier $B$ in the figure. This would be
similar to the Bucher et al.'s nucleation of bubbles by quantum tunneling.
We are working on an adaptation of the work of De Lorenci et al. \cite{DeLor}
to explain this transition, which leads to a de Sitter spacetime with
hyperbolic spatial metric and topology $\mathbf{R}\times (H^{3}/\Gamma
^{\prime }),$ where $H^{3}/\Gamma ^{\prime }$ is a closed hyperbolic
manifold. The scale factor is $R(\tau )=R_{0}\sinh (\tau /R_{0}),$ which
results in substantial inflation over the plateau $C$. We take $%
R_{0}=$ Planck's length.

Finally a phase transition in the true vacuum region $D$ leads to
the radiation era of a Friedmann's spacetime with the same closed topology,
beginning the standard (`big bang') cosmology.

A numerical example was worked out, with $S^{3}/\Gamma $ a lens space $%
L(50,1)$ and $H^{3}/\Gamma ^{\prime }$ Weeks manifold - see \cite{LaLu}. The
present volume of this universe would be about $1/200$ the volume of the
observable space of images - meaning that each source may produce up to 200
images.

\qquad S. S. e C. thanks Funda\c{c}\~{a}o de Amparo \`{a} Pesquisa do Estado
de S\~{a}o Paulo (Brazil) for a doctorate scholarship. H. V. F. thanks
Conselho Nacional de Desenvolvimento Cient\'{i}fico e Tecnol\'{o}gico
(Brazil) for partial financial support.

\vspace{-15pt}
{}




\renewcommand\citep[1]{[\citealt{#1}]}  
\newcommand\citepf[1]{[\citealt*{#1}]}    


\def\.{{\cdot}}
\def\gtapprox{\,\lower.6ex\hbox{$\buildrel >\over \sim$} \, }
\def\ltapprox{\,\lower.6ex\hbox{$\buildrel <\over \sim$} \, }
\def\sun{\odot}
\def\e{ {\scriptstyle \times} 10^}
\def\etal{\mbox{\it et\,al.}}
\def\arcs{\ifmmode {'' }\else $'' $\fi}     
\def\arcm{\ifmmode {' }\else $' $\fi}     
\def\deg{\ifmmode^\circ\else$^\circ$\fi}    
\def\ttimes{{\scriptstyle \times}}
\def\fr7{7$ \hskip -0.9ex \vrule height0.8ex width0.8ex depth-0.73ex
                                                                \hskip0.1ex$}

\def\hMpc{~$h^{-1}$Mpc}
\def\hGpc{~$h^{-1}$Gpc}
\def\frtoday{le\space\number\day\space\ifcase\month\or
  janvier\or f\'evrier\or mars\or avril\or mai\or juin\or
  juillet\or ao\^ut\or septembre\or octobre\or novembre\or d\'ecembre\fi\space \number\year}

\newcommand\joref[5]{#1, #5, {#2, }{#3, } #4}
\newcommand\inpress[5]{#1, #5, #4}
\newcommand\epref[3]{#1, #3, #2}

\def\apj{ApJ}                 
\def\apjs{ApJSupp}                 
\def\aj{AJ}                       
\def\aanda{A\&A}            
\def\cqg{ClassQuantGra}   
\def\mnras{MNRAS}
\def\pasj{PASJ}


\def\affilsize{}
\def\affilsize{\normalsize}
\def\rinj{{r}_{\mbox{\rm \small inj}}}

\subtitle{Observational Methods, Constraints and Candidates}

\subauthor{Boudewijn F. Roukema}
\subaddress{$^1$Inter-University Centre for Astronomy and Astrophysics, \\
 Post Bag 4, Ganeshkhind, Pune, 411 007, India\\
$^2$Institut d'Astrophysique de Paris, 98bis Bd Arago, F-75.014 Paris,
France}

\date{}


%



\submaketitle

\begin{abstract}
All methods of constraining or detecting candidates for the
global topology of the Universe share the same common
principle: an object or a region of space should be
observed several times at, in general, different angular
positions and redshifts. In practice, whether using the
cosmic microwave background (CMB) or collapsed 
astrophysical objects, the practical details of how
an object or a region of space emits electromagnetic 
radiation (degree of local isotropy, evolution) imply
that different strategies have to be adopted, depending on 
which ``standard candles'' are used.

Contrary to popular opinion, it should be noted that the claimed CMB
``constraints'' against small flat multiply connected models are weak:
these are statements about the rarity of simulated perturbation
statistical properties required to match COBE data, rather than about
the consistency of the data with multiple topological imaging on the
surface of last scattering.
\end{abstract}

\subsection{Introduction: a spectrum of differing observational approaches}

Since 1993, much new work in attempting to compare observations
with multiply connected Friedmann-Lema\^{\i}tre models of the Universe
has been carried out. This pioneering work is branching out into 
many different and complementary directions, from cm wavelengths
(CMB) to X-rays, from the Milky Way to 
quasars to galaxy clusters to spots or patches
on the CMB, from close-up investigation of small numbers of objects
to first principles statistical analysis of large would-be 
perfect catalogues, from demonstrations of how
significant detection of cosmic topology would provide constraints
on the curvature parameters to how it would enable measurement
of transversal galaxy velocities. 

It used to be customary to make strong claims 
that ``constraints'' make the small universe idea 
``no longer an interesting cosmological model'', but the renewed
interest in the subject will hopefully lead to more scientifically
worded statements including overt {\em statements of caveats}.

The diversity and vigour of observational cosmic topology is
demonstrated by the fact that at 
this workshop we have a total of nine talks on 
observational approaches (Roukema, Pierre, Wichoski, Uzan,
Weeks, Inoue, Pogosyan, Levin, Bajtlik).
The content of this review itself 
is mostly found in the observational section of \citet{LR99}.
For reviews on cosmic topology in general, see 
\citet{LaLu95,Stark98,Lum98,LR99}. 

\subsubsection{3-D methods}

Marguerite Pierre explained to us how topology can be used to
search for topology. That is, how the 2-D topology of density
contours of hot gas to be detected in X-rays by the XMM 
satellite will represent the local geometry of structure at
redshifts around unity and higher, and can hence be compared
to similar representations of the local geometry in the local
few 100 Mpc in order to find possible 3-D topological isometries
between multiply imaged regions.

Ubi Wichoski took us back to basics. The possibility of identifying a
high redshift image (as a quasar) of our own Galaxy to enough detail
in order to be able to unambiguously prove that it must be an image of
the Galaxy has generally been dismissed as impractical for redshifts
of unity or higher. However, the increasing understanding of the 
Galaxy itself could, in principle, lead to predictions such as the
precise period when the black hole likely to be at the centre
was visible as a quasar. If this were precise enough, then a pair
of opposite quasars occurring at the correct time (and probably in the
direction of the $\rinj$ geodesic) might be sufficient to provide
a convincing candidate 3-manifold.

Statistical methods, either in their most ideal case of an all-sky
complete catalogue of isotropic unevolving emitters or at the other
extreme of finding the few topological image pairs in the haystack of
non-topological pairs, are being further analysed.  Jean-Phillippe
Uzan summarised the French (and Brazilian) work which shows that the
``crystallographic'' (or non-normalised two-point correlation
function) method does not, in general, work for hyperbolic multiply
connected models. This was explained in terms of the different sorts
of pairs which, in the Euclidean case, contribute to spikes in the
histogram.  However, variations on the method such as regrouping all
close pairs in the pair histogram (correlating 
the correlation function) were mentioned and are now
in press \citep{ULL99a}.

\subsubsection{2-D methods}

The optimal two-dimensional (CMB) methods which can lead to 
statements about the consistency or inconsistency of a candidate 3-manifold
and CMB data without making assumptions about the perturbation 
spectrum, methods based on the identified circles principle 
\citep{Corn96,Corn98b}, 
were presented by Jeff Weeks. 

For numerical comparison
of models and observations, Weeks also pointed out some 
convenient mathematical devices for comparing hyperbolic, flat,
and elliptic models, in 2-D for illustration. Use the dot product 
\begin{equation}
\left< (a_x,a_y,a_z), (b_x,b_y,b_z) \right> =
a_x b_x + a_y b_y + a_z b_z 
\end{equation} 
to represent geometrical
operations on the surface $\left< {\bf a}, {\bf a} \right> = 1$, i.e. a sphere 
($S^2$) embedded in $R^3.$ Isometries 
in $S^2$ are represented by unitary real matrices which multiply
by vectors in $R^3$ --- using the dot product.
Then, converting the dot product to
\begin{equation}
\left< (a_x,a_y,a_z), (b_x,b_y,b_z) \right> =
a_x b_x + a_y b_y - a_z b_z 
\end{equation} 
gives 3-D Minkowski space, 
i.e. like $R^3$ but with the implied metric from the new dot product.
The surface $\left< {\bf a}, {\bf a} \right> = 1$ is now a hyperbolic
surface, $H^2,$ instead of a sphere, and isometries are represented
by matrices whose component vectors are orthonormal under the
new dot product. This of course generalises to the 3-D case.

Although the perturbation simulation approach to exploring CMB data
has so far been used to make statements about perturbation statistics
rather than directly about topology, the approach is still useful
and challenging computationally and mathematically.
Kaiki Taro Inoue demonstrated calculation of eigenmodes 
in compact hyperbolic universes, which have previously been considered
as exceedingly difficult to calculate.

Dmitri Pogosyan reminded us of the very thorough CMB simulations for two
hyperbolic models carried out by himself and his collaborators. 
Janna Levin reviewed her and her collaborators' 
simulational work relating to horn topologies
and on ideas for pattern searching for spots in the CMB as an alternative
to the identified circles method and the perturbation simulation methods.

\subsubsection{Consequences}

The consequences of multiple topological imaging are not merely
secondary questions which can lay in wait for a discovery to be 
considered significant. If a correct detection is made, it should
help ``fit pieces in a puzzle''. 
Stanislaw Bajtlik pointed out how multiple topological images of
galaxy clusters could enable estimation of galaxy velocities 
transversal to the line-of-sight. Given a moderate scale
photometric and spectroscopic programme on a good telescope,
this should tighten understanding
of dynamics of clusters, which would in turn relate to dynamical
estimates of the curvature parameters ($\Omega_0, \lambda_0$),
which ought to themselves be consistent with the claimed topological
detection. Such self-consistent loops could enable a considerable
range of different physical arguments to be sharpened up so 
that 10\% would no longer be considered a high precision 
for observational estimates of cosmological parameters.

\subsection{Comparison of different approaches}

Although the different methods are given different names, they
all share the same principle: an astrophysical collapsed object
or a region of plasma at the epoch of last scattering has to be
viewed multiply in different directions in order to reveal
the multiple connectedness of the Universe. The object or region
of plasma should ideally be a ``standard candle'' in order for
a search or an attempt to refute a candidate 3-manifold or 
a set of candidate 3-manifolds to give a result with a minimum
of caveats. 

The differences between the approaches then divide into 
\begin{list}{(\alph{enumi})}{\usecounter{enumi}}
\item the choice of which standard candles to use (e.g. those with
a 3-D or a 2-D spatial distribution),
\item the means of compensating for the observational difficulties
(i) to (vi) [\S5.3 \citealt{LR99}; plus (vii) gravitational lensing]
for those particular standard candles
\item the choice of whether 
  \begin{list}{(\alph{enumi}-\roman{enumii})}{\usecounter{enumii}}
  \item to test self-consistency of positions of
  known objects or plasma regions with 3-manifolds or
  \item to simulate structure in the Universe for given 3-manifolds 
  and estimate the probability
  that the statistical properties of the observed structures could have
  been drawn from distributions of those same properties for the simulated
  structures.
  \end{list}
\end{list}

A brief summary of the more recent choices for (a) and (b) are
listed in Table~\ref{t-obsvn}, and are discussed to some extent
in \S5 of \citet{LR99}. The option (c-ii) has (to the best knowledge
of the author) only been applied to COBE data, not to 3-D data,
and, apart from \citet{Rouk99,Rouk00c} in which (c-i) is applied, 
is the {\em only} alternative which has
so far been applied to COBE data.

Given that the assumptions generally made about structure in the
Universe, i.e. assumptions
about statistics of the perturbation spectrum, 
are based on inflationary theory which is unlikely to predict, for
example, a flat multiply connected universe of less than the horizon
size, it can be expected that these assumptions fail at some level
on the length scales approaching that of the Universe. That is,
the assumptions about structure are unjustified theoretically
at scales $L$ where $L \ltapprox \rinj < r_+.$ (See \citet{LR99} 
for definitions of $\rinj,$ $r_+.$) 

They are equally unjustified
observationally: the only observations known to reliably describe
structure on super-Gigaparsec scales are those of COBE --- analysed
under the assumption of simple connectedness. The COBE data could, 
of course, be reanalysed under a multiply connected hypothesis,
and the properties of the perturbations required in order to fit
the model could be presented. This would be a useful project 
to carry out, and might result in a long list of candidate 
multiply connected, small ($2\rinj \sim 1${\hGpc}?), flat models 
which are consistent with COBE data...

\begin{table*}[htb]
\caption[Summary of methods and observational results.]{
\label{t-obsvn} Summary of the most recent methods
and observational results, adapted from
\S5 of \protect\citet{LR99}.
Author abbreviations are 
LLL96 (\protect\citet{LLL96}), 
RE97  (\protect\citet{RE97}), 
R96   (\protect\citet{Rouk96}), 
CSS96/98 (\protect\citet{Corn96,Corn98b}), 
BPS98  (\protect\citet{BPS98}), 
RB99   (\protect\citet{RBa99}). }

\begin{tabular}{cc | c c c }
\hline
\multicolumn{2}{l}{{\bf (1) Methods:}} & g. clusters & QSO's & CMB \\
\multicolumn{1}{l}{3D:} \\
 clus opt & cosmic crystallog. & LLL 96 \\
 clus Xray & brightest cluster & RE 97 \\
 QSO's& local isom. search & & R 96 & \\
\hline
\multicolumn{1}{l}{2D:} \\
 CMB & ID'd circles & & & CSS96/98 \\
     & $C_l$ --- cutoff & & & many \\
     & correlation fn & & & BPS98 \\
\hline
\multicolumn{3}{l}{{\em Ideal object:}} \\

& no Evoln & monotonic E & strong E & weak E? \\
& zero pec velocity& prob. small & prob. small & N \\
& isotropic emitter& Y (nearly) & N & Y/N \\
& seen to large $z$& Y ($\kappa_0 < 0$), N (o.w.) & Y & Y \\
& seen over large vol& N & Y & Y (sph shell) \\
& seen to $|b^{II}| \ll 20\deg$ & N & N & N \\
& no g. lensing &  OK & few \arcs & ? \\

\multicolumn{3}{l}{{\em Assumptions on $\kappa_0$, $\{g_i\}$, ideal=
none:}}  \\
&&     none & $\kappa_0$ (use range) & circles: none \\
&& & &                                 $C_l$: all \\
\hline \\
\multicolumn{3}{l}{{\bf (2) Constraints:}}\\
& & CC: $2r_+ \gtapprox R_H/20$ \\
& & BC: $2r_+ \gtapprox R_H/10$ \\
& & & N/A \\
\multicolumn{5}{l}{For the following special cases, but really testing
perturbation spectrum assumptions:}\\
\multicolumn{5}{l}{$\kappa_0=0,$ if $\theta(g_i,g_j)= 90i\deg$ or
$60i\deg, i \in Z$ then}\\
&&&& ($2\rinj \gtapprox R_H/2$) \\
\multicolumn{5}{l}{$\kappa_0<0,$ if $\Gamma=m004(-5,1)$ or
$\Gamma=v3543(2,3)$ then} \\
&&&& ($2\rinj \gtapprox 2 R_H$) \\
\hline \\
\multicolumn{3}{l}{{\bf (3) Specific candidates:}}\\
&& serendipitous & 2$\sigma$ implicit & ``preferable\\
&&&& to SCDM''\\
&& $\kappa_0=0$ & $\kappa_0<0$? & $\kappa_0=-0.2$ \\
&& $\widetilde{M}/\Gamma= T^2 \times {\bf R}$ & [non-orientable] &
v3543(2,3) \\
&& $2\rinj$ ($\Omega_0$) = && $ \rinj =0.95 R_H$\\
&&$965\pm5h^{-1}$Mpc ($1$)& \\
&&$1190\pm10h^{-1}$Mpc ($0.2$) &\\
&&RE97, RB99& R96 & BPS98 \\
\hline
\end{tabular}
\end{table*}

\subsection{Candidates versus constraints}

The history of observational cosmology shows that strong claims
can be made which are mutually inconsistent, and that systematic
errors are often understated or missed entirely.

A challenge for testing the solidity of the claimed constraints on
the values of $\rinj$ and $r_+$ is to attempt to correctly 
refute specific candidate 
3-manifolds [e.g. part (3) of Table~\ref{t-obsvn}],
taking into account all the assumptions and analysing 
the possibilities that the assumptions may be wrong.
This may help convince 
telescope time committees that a thorough observational attitude is
being taken to cosmic topology.

\subsection{Conclusion and suggestions for the future}

The rapidly increasing amount of data on scales of $1-20h^{-1}$~Gpc,
i.e. $\sim (0.1-2)R_H$\footnote{horizon radius; $2R_H$ 
is the horizon diameter}, the combination of
advantages and disadvantages of different objects or emitting regions
and the diversity of possible analysis strategies imply that
creativity and care in modifying or combining the different approaches
are likely to be necessary in order to obtain a significant detection
of --- or a significant $R_H$ scale constraint against --- the
topology of the Universe.

The history of observational cosmology tells us that ``tricks'' which
may not even by theoretically understood may be the key to making
simple principles applicable. 

For example, the supernova Ia calibration method
which makes SNe-Ia a better standard candle than before is essentially
an empirical technique, but is giving impressive results about the curvature
parameters ($\Omega_0$ and $\lambda_0$) based on the classical
apparent magnitude--redshift relation 
\citep{SCP9812}, which otherwise was
considered too inaccurate to apply in practice to real astrophysical
objects.

{\em What ``tricks'' are possible to step around or
correct for the various problems listed in Table~\ref{t-obsvn}?}

A related strategy for estimating the curvature parameters 
is the combination of SNe-Ia and COBE data, which
give ``orthogonal'' constraints on the relation between the two
curvature parameters. 

{\em Could an analogy of this idea be useful in cosmic topology?}

Apart from analysis of new data sets, answers to these questions 
may help extract information which is present but hidden in 
existing data... Rendez-vous at the next workshop.





\subtitle{Topological Images of the Galaxy}

\subauthor{U. F. Wichoski}
\subaddress{Department of Physics, Box 1843, Brown University, \\
Providence, RI 02912, USA\\
{\normalsize Present address: Depto. 
de F{\'\i}sica, IST-CENTRA,}\\ 
{\normalsize Av. Rovisco Pais, 1 - 1096 Lisboa Codex, 
Portugal.}\\
{\normalsize \em E-mail: wichoski@x9.fisica.ist.utl.pt}
}

\submaketitle

\begin{abstract}
\noindent 
One of the possibilities to constrain the topology of the Universe 
to the observational data is to search for topological images of 
our own Galaxy. This method is based on the idea that in a 
multi-connected Universe we would, in principle, be able to 
see the light emitted by our own Galaxy in the early stages 
of its evolution. The significant 
identification of these images would give 
us strong evidence that the topology of the Universe 
may be non-trivial. 
\end{abstract}


\subsection{Introduction}

In the standard big-bang model the Universe is described 
by a spatially homogeneous isotropic Friedmann-Lemaitre model. 
Mathematically this model is represented by a 
4-dimensional manifold $M$ endowed with a Lorentzian metric 
$g_{ab}$ such that the requirement that $(M,g)$ is stably causal 
is fulfilled 
(for terminology and mathematical definitions 
we refer the reader to the excellent review by M. Lachi\`eze-Rey and 
J.-P. Luminet \cite{lalu} and references therein, and for an 
update \cite{lumi} and \cite{luro}). 

The homogeneity and isotropy imply that the curvature of 
the 3-manifold that describes the spatial section $S$ 
of the spacetime 4-manifold $(M,g) = (S \times T,g)$ 
is constant. 
The spatial curvature can be parameterized 
by a constant $k = 1, 0, -1$ describing the  
negative, zero and positive cases respectively. 
In terms of the Robertson-Walker metric \cite{wei}  
\[   
ds^2 = c^2 dt^2 - R^2(t) \{ \frac{dr^2}{1 - k r^2} + r^2 d\theta ^2 + 
r^2 sin^2\theta d\phi ^2 \} \;,
\]
\noindent
where $R(t)$ is the scale factor.

The determination of the curvature of the Universe is still an 
open question, and, in principle it depends on the 
determination of the cosmological 
parameters $\Omega_0$ and $ \Omega_{\Lambda}$ by 
the observations \cite{tu98} 
(see \cite{rolu} for a new method by which $\Omega_0$ and 
$\Omega_\Lambda$ can be precisely estimated). 

The determination of the curvature, however, is related to the 
local properties of the spacetime, i.e., 
to the metric. It has been usually taken 
for granted that the global properties of the spatial section of 
the spacetime are those of a simply-connected 
3-manifold: The infinite hyperbolic space $H^3$, the infinite 
Euclidean plane $E^3$, and the hypersphere $S^3$. The spacetime 
manifold is then represented by

\begin{itemize}

\item $H^3 \times T \rightarrow$ in the case of 
hyperbolic spatial section of negative constant curvature; 

\item $E^3 \times T \rightarrow$ in the case of 
Euclidean spatial section of null curvature; 

\item $S^3 \times T \rightarrow$ in the case of 
spherical spatial section of positive curvature. 

\end{itemize}
\noindent

This assumption implies that the Universe in 
the case of negative and zero curvature is spatially infinite. 
No direct evidence that this assumption 
is correct has been found. 

If we drop the supposition that the spatial section of the 
Universe is simply-connected, 
we allow for the possibility that the spatial sections are 
multi-connected 3-manifolds. These spaces are compact 
in at least one spatial dimension and their volume can be finite 
(in the case it is compact in all three dimensions) 
irrespective of the value of the curvature. 
There is an infinite number of topological classes related to 
multi-connected 3-manifolds \cite{lalu}.

A topological class is characterized by 
the fact that any compact 
3-manifold $M$ of constant curvature $k$ can be expressed as 
the quotient space $M = {\tilde{M} \over \Gamma}$, where $\tilde{M}$ 
is the universal covering space 
($\tilde{M} = H^3, E^3, S^3$ for $k = -1, 0, 1$ 
respectively) and $\Gamma$ is a subgroup of 
isometries of $\tilde{M}$ acting freely and discontinuously 
\cite{lalu}. 
This implies that the multi-connected 3-manifold can be divided 
into simply-connected domains (a tessellation of $\tilde{M}$) 
any of which can be considered to be the 
so-called fundamental polyhedron or fundamental cell. 

The consequence of the multi-connectedness is that there would 
exist more than one geodesic linking a source to the observer 
which implies that an astronomical object can have  
topological copies of itself. 
These multiple images would be, in principle, visible 
simultaneously to an observer at a given time 
even for the cases of constant 
curvature negative and zero (it is always possible in the case 
of constant positive curvature, for $\Omega_\Lambda$ 
values for which the Universe is old 
enough, because $S^3$ is compact). 
Nonetheless, they would, in general, 
correspond to the object seen at different lookback times. 
Apart from astrophysical conditions (see below), the only 
requirement for the observation of the topological images 
is that the size of the Universe must be smaller 
than the horizon. 

The observable Universe is a subset (interior of a sphere) of 
the universal covering space, 
ie, the locus of the images of the fundamental polyhedron 
within the horizon diameter. 
There is one and only 
one topological image of an object in each cell in which the 
universal covering space is tessellated. 
The images of the astronomical 
objects lying inside the cell to which the observer 
is placed are all considered in this paper 
to be the real images of the 
object. The other images of the object (lying in adjacent 
cells) are considered here to be topological or ghost images 
(note that we are not considering the case 
of gravitational lensing).

\subsection{The search for topological images}

To determine in which kind of Universe we live, besides the standard 
cosmological parameters, we need an extra set of topological 
parameters: the specification of the 
base manifold $\tilde{M}$ and its subgroup of 
isometries $\Gamma$. In a way more suitable for the observations 
we instead  characterize 
the size of the fundamental polyhedron by defining 
\cite{cornetal} the injectivity radius, $r_{ij}$, as 
half of the smallest distance from an object to 
one of its topological images; and the out-radius 
$r_{+}$, the radius of the smallest sphere in the 
covering space which totally includes the fundamental 
polyhedron.

As we have mentioned before, in a multi-connected Universe, if 
the size of the horizon diameter 
is smaller than the size of the 
Universe, it would be, in principle, possible to observe 
topological images of an object.  
 
It is reasonably well established that there is a lower limit 
in the size of the Universe $r_{inj} > 100$ Mpc \cite{luro}. 
The interest is in probing scales from that size up to the horizon.

Based on 3-D methods from the previous work of 
Demianski and Lapucha \cite{dela}, 
Fagundes and 
Wichoski \cite{fawi}, Roukema \cite{ro}, and 
Roukema and Blanloeil \cite{robla} we can 
draw the general characteristics of this kind of 
search. 
Ideally, it would be possible to determine the topology 
of the Universe by performing the following method:

\begin{enumerate}

\item One of the three possible curvatures of 
the spatial section of the spacetime 
manifold (negative, null, or positive) 
is chosen based on theoretical 
reasons and/or observational data or simplicity; 

\item Next a topological class 
must be chosen, again based on either theoretical reasons and/or 
observational data or simplicity; 

\item An astronomical object (galaxy, cluster of 
galaxies, supercluster of galaxies, quasar, or a 
cosmic microwave background photon emitting region) 
must be chosen; 

\item  Calculations using the subgroup of isometries $\Gamma$ associated to 
the topology chosen applied to the chosen object (or objects) give 
a pattern in the sky to be searched (this step may be skipped 
see \cite{ro}); 

\item Search for the topological images of the object(s). 

\end{enumerate}
There are of course many difficulties in applying this 
method: 

\begin{enumerate}
\item The criterion for choosing the geometry can be 
the recent data (e.g. high $z$ SNe-Ia \cite{supernova}), but 
the criterion for choosing the topology is much weaker and 
our present knowledge of the Universe allows us only to say
that some topological classes are disfavored. 
If simplicity is chosen we may incur the same 
oversimplification that has led us to the simply-connected 
topologies. 

\item What kind of astronomical object would be more suitable to 
be seen as a topological image? When choosing a 
determined class of objects we have to bear in mind that it is 
important to understand 
the evolution of this class of objects. The recognition 
of these objects in their earlier stages of 
evolution depends on that. 

\item The topological images may not all be visible: 
\begin{itemize}
\item The more distant the topological image, the weaker 
its luminosity;
\item the emission can be non-isotropic (the object could have 
different appearance when observed by different angles);
\item regions of the intergalactic medium (IGM) can absorb or 
scatter the light in its way to the observer;
\item topological images can be hidden behind an 
astronomical object;
\item distortions of the topological images due to gravitational 
lensing could prevent us from recognizing them; 
\item peculiar velocities can shift the topological images (this 
is critical only if we are trying to fit a specific topological 
model).
\end{itemize} 
\end{enumerate}

\subsubsection*{Searching for topological images of the Milky-Way}

Due to so many difficulties to fit a topological model 
we can think of limiting the scope of the search program. The 
basic idea is to observe a topological image of an object 
without a particular topological model in mind. 

In principle, because we have an insider's view of our Galaxy 
we are in an advantageous position to understand its evolution 
and peculiar characteristics.
It would allow us to recognize 
the Milky-Way in earlier stages of its evolution as a 
topological image of itself. Because of our privileged 
position in relation to the topological images of the 
Milky-Way (for every high $z$ image of the Milky-Way, 
another image should exist at the same $z$ irrespective 
of topology, and in many cases the second image 
will be antipodal to the first; see \cite{fawi}) 
the observation of some of them 
may be attainable before a full determination 
of a topological model by other techniques. 
Other techniques for determining global topology could 
profit from these observations and start with a 
reduced parameter space. Given the enormous 
numbers of comparisons of data points required for methods 
which avoid prior assumption of topology, this would be a 
a considerable advantage 
and would make it easier to fit a topological model. 
In this respect it is important to 
remark that the recognition of just one topological image 
of the Milky-Way may be enough to convince ourselves that we 
live in a Universe with multi-connected topology and 
with at least 
one compact spatial dimension. 
The search for topological images of the Milky-Way may be 
not effective to determine the topological class or even 
the signal of the curvature. The scope of the search can  
be limited to the search of a single, if any,  
visible topological image of the Milky-Way. 
Complementary methods 
using the topological images of the Milky-Way found by 
the present method, as 
we have mentioned above, can eventually take over 
the determination of the topological model.  

Considering the fact that galaxies are now 
found with redshifts as large or maybe 
larger than those of the quasars 
makes it pertinent to ask if we should look for topological 
images of our Galaxy as a quasar or as a high redshift galaxy. 
The answer again is in the understanding of the evolution 
of our Galaxy.   
The Sloan Digital Sky Survey (SDSS) and new satellites as 
X-ray Multiple Mission (XMM) will provide more and better 
data, this way improving the chances of detecting quasars which 
could be the topological images of the center of our Galaxy, 
followed up by optical/NIR imaging on large telescopes (VLT, Keck) 
to see if the predecessors of the present thin disk, thick disk, 
bulge, bar and halo components of the Galaxy and the surrounding 
Local Group galaxies are present around those same quasars.

We would like to stress here that we can consider 
the search for the topological 
images of the Galaxy as an intermediate method in the 
determination of the topology of the Universe.  

\subsection*{Acknowledgments}
This work was supported (at Brown) in part by the US Department 
of Energy under contract DE-FG0291ER40688, Task A. 
I would like to thank CTP98's Organizing Committee for the 
invitation and financial support provided; LIP and IST-CENTRA for 
financial support; Ana Mour\~ao for the warm reception in Lisbon; 
and Boud Roukema for his invaluable comments and suggestions.




\newcommand{\del}{\partial}
\newcommand{\m}{\mathbf}
\newcommand{\x}{{\mathbf{x}}}
\newcommand{\y}{{\mathbf{y}}}
\newcommand{\z}{{\mathbf{z}}}
\newcommand{\n}{{\mathbf{n}}}
\newcommand{\U}{\underline}
\newcommand{\f}{\frac}
\newcommand{\T}{\tilde}
\newcommand{\N}{\nonumber}
\newcommand{\bb}{\bibitem}
\newcommand{\BF}{\begin{figure}}
\newcommand{\EF}{\end{figure}}
\newcommand{\BE}{\begin{equation}}
\newcommand{\EE}{\end{equation}}
\newcommand{\BEA}{\begin{eqnarray}}
\newcommand{\EEA}{\end{eqnarray}}
\subtitle{Comments on the constraints on the topology of compact
low-density universes}
\subauthor{Kaiki Taro Inoue}
\subaddress{Yukawa Institute for Theoretical Physics, Kyoto University\\
Kyoto 606-8502, Japan}
\submaketitle

\begin{abstract}
Although it has been argued that ``small universes'' being multiply
connected on scales smaller than the particle horizon are ruled out,
it is found that constraints on compact multiply connected models 
with low matter-density (flat or hyperbolic) are not stringent.
Furthermore, compact hyperbolic models ($\Omega_0=0.1\sim0.2$) with
volume comparable to the cube of the
present curvature radius are much favored compared to the
infinite counterparts due to the mild suppression on large-angle power.
\end{abstract}

\section*{}
In the framework of modern cosmology, one often takes it for granted that
the spatial geometry of the universe with finite volume is limited to 
that of a 3-sphere. However, if we assume that the spatial
hypersurface is multiply connected, then the spatial geometry of 
finite models can be
flat or hyperbolic as well. It should be emphasized that 
``open'' models can be closed
by a certain set of appropriate identification maps, 
which have long been ignored by cosmologists. 
\\
\indent
Since 1993,  a number of articles concerned with constraints
on the topology of flat models with no cosmological
constant using COBE-DMR
data have been 
 published(Sokolov 1993; Stevens, Scott \& Silk 1993; de Oliveira, Smoot \& 
Starobinsky 1996; Levin, Scannapieco \& Silk 1998). 
The large-angle temperature 
fluctuations discovered by the COBE satellite constrain the 
topological identification scale $L$ (twice the injectivity radius) 
larger than 0.4 times the diameter of the observable region 
4$H_0^{-1}$;in other words, the
maximum expected number of copies of the fundamental domain inside 
the last scattering surface is $\!\sim$8 for compact flat
models without the cosmological constant\footnote{The constraints are
for models in which the diameter of the space is comparable to twice
the injectivity radius $R_{\textrm{inj}}$. However, if the diameter is much longer
than $R_{\textrm{inj}}$, 
then the constraints are not so stringent(Roukema 2000).}.
\\
\indent
Fluctuations on scales larger than 
the diameter of the spatial hypersurface in every direction 
are strongly suppressed although skewed fluctuations 
can have large correlation length in some directions. 
If one assumes the usual
Harrison-Zeldovich type initial power, then the large-angle
power is strongly suppressed for small compact models 
without the cosmological constant.
\\
\indent
In contrast, a large amount of large-angle fluctuations can be
produced for compact low density models due to the decay of 
gravitational potential near the present epoch which is known as the 
integral Sachs-Wolfe effect(Cornish,
 Spergel \& Starkman 1998). If the spatial geometry is
sufficiently flat or hyperbolic then the physical distance of two separated 
points which subtends a fixed angle at the observation point 
becomes larger as the points are
put at distant places. Large-angle fluctuations can be
generated at late epoch when the fluctuation scale ``enters''
the topological identification scale $L$. Recent 
statistical analyses using only the power spectrum have shown that 
the constraints on the topology are not stringent for small
compact hyperbolic (CH) models including the smallest (Weeks) and the
second smallest (Thurston) known manifolds and an non-arithmetic orbifold
(Cornish \& Spergel 2000, Inoue 2000a, Aurich 1999).
\\
\indent
These results are clearly at odds with the previous constraints
(Bond, Pogosyan \& Souradeep 1998,2000) on CH models
based on pixel-pixel correlation statistics.
They claim that the statistical analysis using only 
the power spectrum is not sufficient since it
 can describe only
isotropic (statistically spherically symmetric) correlations.
This is true inasmuch one considers fluctuations observed from 
a particular point. Because any CH manifolds are globally
anisotropic, expected fluctuations would be statistically 
globally anisotropic at
a particular point. 
\\
\indent
In order to constrain CH models, it is necessary to 
compare the expected fluctuation patterns 
observed from any place for all the possible orientations 
of the observer to the data
since CH manifolds are also globally 
inhomogeneous. It should be emphasized that the 
constraints obtained in the previous analyses 
are only for CH models
at a particular observation point $Q$ where the injectivity radius is locally maximum 
for 24 particular orientations. The point $Q$ is
rather special one in the sense that some of the mode functions
(eigenfunctions of the Laplacian) have a symmetric structure.  
It is often the case that the base point $Q$ becomes a fixed point of
symmetries of the Dirichlet domain or the manifold.
\\
\indent
In order to see the dependence of the likelihood 
on the position and the orientation of the observer,  
temperature correlations in
the Thurston models $\Omega_0=0.2$ and $\Omega_0=0.4$ 
without the cosmological constant are compared 
with one realization of fluctuations in the Einstein-de-Sitter model with
an angular power 
$C_l\propto 1/(l(l+1))$ at 24 pixels (resolution 2).
We use the pixel-pixel based Bayesian
likelihood analysis for testing these models.
To reduce the computation time,
we cut all multipoles $l>5$ as well as monopole and dipole.  
In this work, the first 36 eigenmodes obtained by
the direct boundary element method were used. 
If correlations due to the non-trivial topology
are irreconcilable to the COBE-DMR data, the likelihoods
will be considerably worse than those using only the angular
power spectra.
\\
\indent
\begin{figure*}[t]
\centerline{\psfig{figure=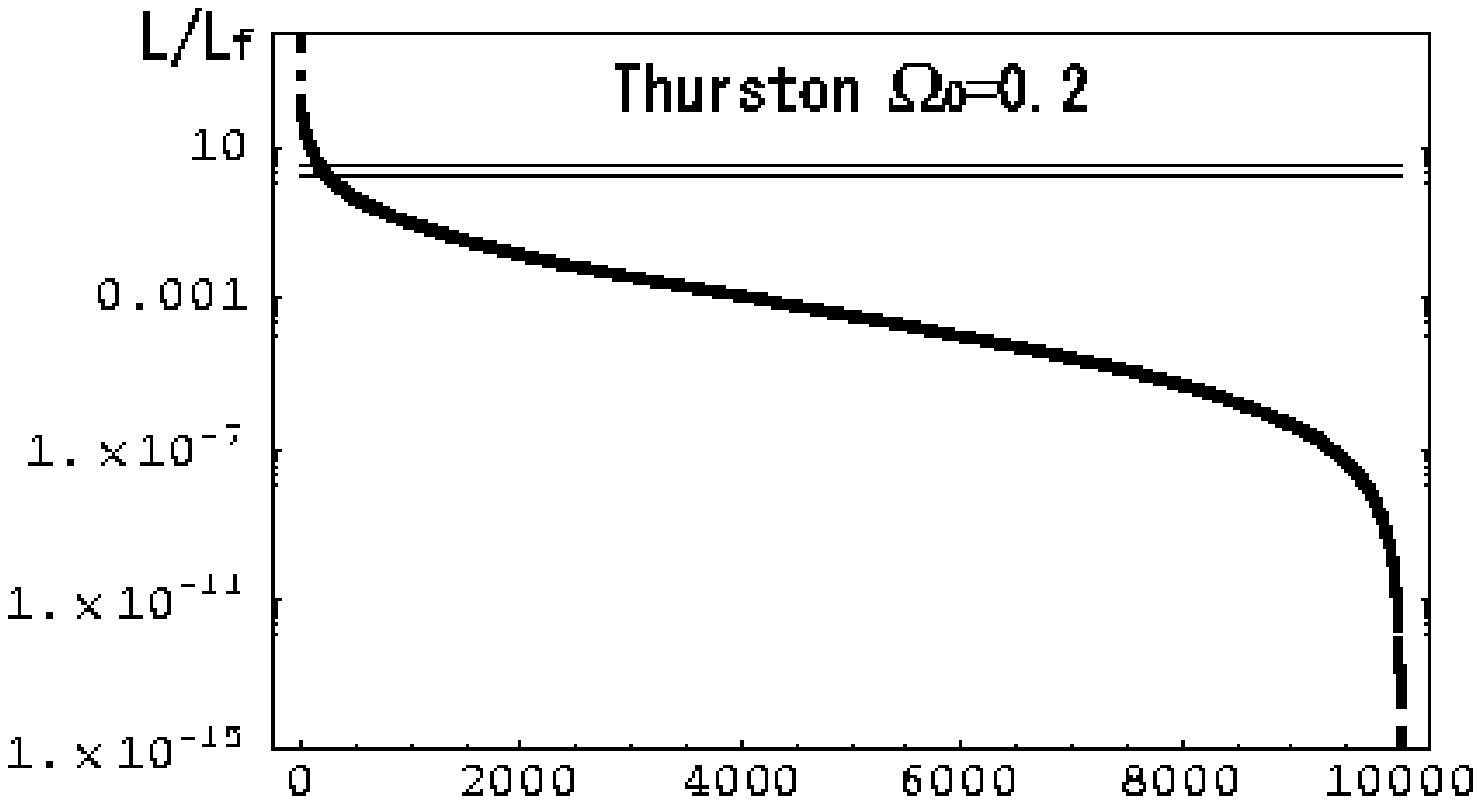,width=8.5cm} 
\psfig{figure=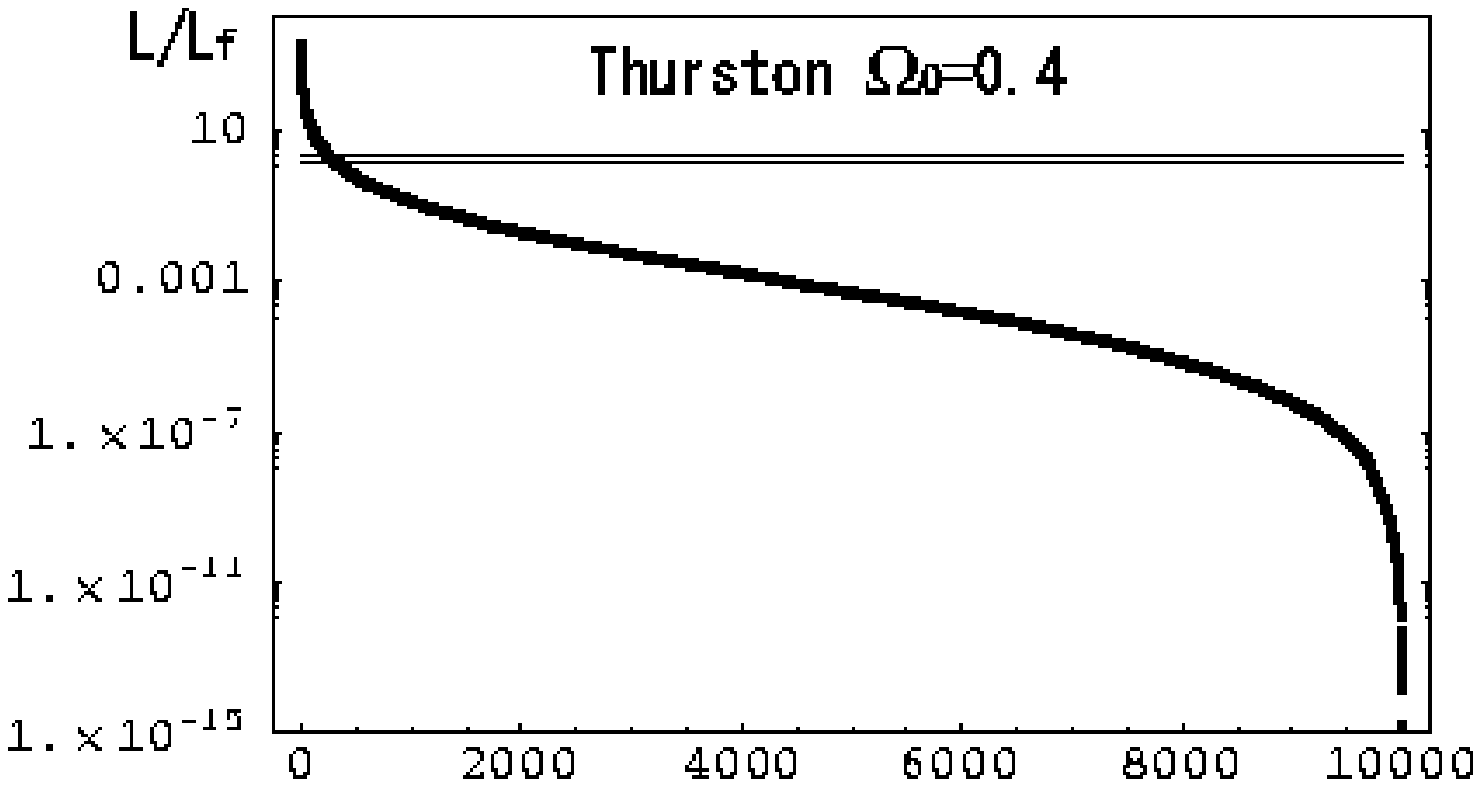,width=8.5cm}}
\caption[Plots of likelihoods of Thurston models relative 
to Einstein-de Sitter models.]{Plots of likelihoods  (marginalized over
normalization $Q=\sqrt{5 C_2/(4 \pi)}$)
in descending order for the Thurston models 
($\Omega_0=0.2,0.4$)  ${\cal{L}}$ over that for the simply connected
Einstein-de Sitter model ${\cal{L}}_f$ 
for each one of the total of 10000 realizations (100 positions and 
100 orientations of the observer). The toy map is produced by
one realization in the simply connected 
Einstein-de Sitter model($C_l \propto
1/(l(l+1))$) that is similar to the COBE data on large-angle scales. 
Initial fluctuations are
assumed to be Gaussian. All multipoles except for $2\le l \le 5$
are ignored. In each plot, the upper and the lower 
horizontal lines denote the 
value of the likelihood in which anisotropic components in the 
two-point correlation are neglected(Gaussian approximation), 
the value of the averaged 
likelihood in which all components are included(``rigorous method''), 
respectively.}
\label{fig:EMW}
\end{figure*}
However, as shown in figure 1, 
the obtained likelihoods marginalized over the 
normalization ($Q=\sqrt{5 C_2/(4 \pi)}$) in the power, the
position and the orientation of the observer are found to be 
comparable to that using
only the power spectrum.
The distribution of the likelihood function has some peaks
at a particular choice of position and 
orientation of the observer. The
likelihoods
are dominated by only 2 to 3  percent of the 10000 realizations.
As was pointed out by Bond et al, the predicted correlation 
patterns are preferred over the infinite counter part 
for some specific choices while 
most choices of the position
and orientation are ruled out. 
\\
\indent
The result is not surprising if one takes the pseudo-random
behavior of the mode functions into account (Inoue 2000b).
Each choice of position and orientation of the
observer corresponds to a ``realization'' of independent random 
Gaussian numbers. By taking an average over the position and the
orientation, a set of anisotropic patterns all over the place in the CH
space comprises an almost isotropic random field. Consider two realizations 
$A$ and $B$ of such an isotropic random field. The chance you would
get an almost similar fluctuation pattern for $A$ and $B$ would be very low but
we do have such an occasion. Similarly, likelihoods at some
particular position and orientation are usually very low
but there are cases where the likelihoods are considerably high.
\\
\indent
Assuming that the initial perturbations are also Gaussian, then
the expected fluctuations are described by an isotropic non-Gaussian random
field.  The distribution functions of the expansion coefficients of
the fluctuations on the sky have a peak at the expectation value ($=0$)
and decrease slowly in the large limit compared with the Gaussian one with
the same variance since the fluctuations are written in terms of the
sum of products of two independent Gaussian variables determined by
the initial condition and the geometry of the 
background space, respectively (Inoue 2000b). The cosmic variances at large
angle-scales are somewhat
larger than that for the Gaussian models since the effect of
the ``geometric'' variance due to the global inhomogeneity of the
background space. 
\\
\indent
On the other hand, if we assume a uniform
prior probability for the initial fluctuations having the same
wave-number dependence as the extended Harrison-Zeldovich
spectrum, then the fluctuations can be well described as isotropic
Gaussian fields owing to the pseudo-Gaussianity in the mode functions.
Note that the power spectrum completely specifies the
correlation structure for any Gaussian models. 
If there is no nearly perfect alignment
to the expected values in the data,
we expect that the statistical tests using 
this Gaussian approximation
give lower bounds for the likelihood since the cosmic variance
takes the minimum value for an isotropic Gaussian field. Thus
the constraints obtained using only the
power spectrum can be verified. 
\\
\indent
Bayesian analyses for the Weeks and the Thurston models
with or without the cosmological constant have been done
using the inverse-noise-variance-weighted
average map of the 53A,53B,90A and 90B COBE-DMR channels(Inoue 2000c).
In the analyses, the Gaussian approximation was used.
Surprisingly, it is found that these models are much favored
than the infinite counterparts for $\Omega_0=0.1\sim0.2$.
This is because the 
excess of large-angle power due to the integrated Sachs-Wolfe
effect is reduced owing to the mode cut-off.
\\
\indent
What about compact flat models with the cosmological
constant? From the full Bayesian analyses it is found that the 
possible number of the 
copies of the fundamental domain inside the observable region
is 50-60 for a flat 3-torus model with $\Omega_\Lambda=0.9$ 
(Inoue 2000c).
In contrast to CH manifolds, flat 3-torus is
globally homogeneous and the fluctuations cannot be statistically
isotropic. The large-angle suppression 
is not stringent but the angular powers have a jagged structure
owing to the global anisotropy. Even though, the small signal-to-noise
ratio in the COBE data on small angular scales $l>15$ makes
it difficult to determine whether a prominent 
jagged power is observationally 
allowed or not.
\\
\section*{Acknowledgments}
I would like to thank N. Sugiyama and T. Chiba for their useful
comments and A.J. Bandy for his advice on the use of the COBE
data. I would also like to thank J. Weeks for sharing his expertise
in the topology of 3-manifolds and B. F. Roukema for 
providing me the opportunity for submitting this short contribution.
K.T. Inoue is supported by JSPS Research Fellowships 
for Young Scientists, and this work is supported partially by 
Grant-in-Aid for Scientific Research Fund (No.9809834).


\newcommand{\be}{\begin{equation}}
\newcommand{\ba}{\begin{eqnarray}}
\newcommand{\ee}{\end{equation}}
\newcommand{\ea}{\end{eqnarray}}

\subtitle{Topological Pattern Formation}

\subauthor{Janna Levin and Imogen Heard}
\subaddress{Astronomy Centre, University of Sussex\\
Brighton BN1 9QJ, UK\\
{\em email: janna@astr.cpes.susx.ac.uk}}


\submaketitle

\begin{abstract}
We provide an informal discussion of pattern formation in a finite
universe.
The global size and shape of the universe is revealed in the pattern
of hot and cold spots in the cosmic microwave background.
Topological pattern formation can be used to reconstruct the geometry
of space, just as gravitational lensing is used to reconstruct the
geometry of a lens.
\end{abstract}


\bigskip
\medskip


\bigskip
\medskip


We have all come to accept that spacetime is curved.
Yet the idea that space is topologically connected still meets with
resistance.
One is no more exotic than the other.
In the true spirit of Einstein's revolution, 
gravity is a theory of geometry and geometry has two 
facets: curvature {\it and} topology.  

The big bang paradigm forces us to consider the topology of the universe.
As best as we can ascertain, when the universe 
was created both gravity 
and quantum mechanics were at work.  
Any theory which incorporates gravity and quantum mechanics must
assign a topology to the universe.
String theory is currently the most powerful model which naturally
hosts gravity in a unified framework.
It should not be overlooked that in string theory there are six extra 
dimensions all of which must be topologically compact.
In order to create a viable low-energy theory, the internal dimensions 
are finite Calabi-Yau manifolds.
We naturally wonder why
a universe would be created with
six compact dimensions and four infinite ones.
A more equitable beginning might create all spatial dimensions compact
and of comparable size.  Six dynamically squeeze down while the
other
three inflate.  In fact, it is dynamically possible for inflation of 
$3$-space to
be kinetically driven by the contraction of internal dimensions
\cite{jin}.
Whatever mechanism stabilizes the internal dimensions at a small size
would likewise stabilize the external dimensions at an inversely large
size.
Topology need not be at odds with inflation.

Another interesting possibility is that the topology itself naturally
selects the expansion of $3$-dimensions and the contraction of $6$.  The
topology can create boundary contributions to an effective
cosmological constant.  The sign and magnitude of the vacuum energy
depends on the topology and it is 
conceivable that it selects three dimensions for expansion and three
for contraction in a kind of inside/out inflation.  
In the wake of the recent observational evidence
that there is a cosmological constant today, the pursuit of 
these calculations is worthwhile.  Perhaps 
we are still inflating as the vacuum energy tracks the topology
scale.

Our quest to measure the large-scale curvature of the universe may
also produce a measurement of the topology.
(For a review and a collection of papers see \cite{{lum},{volume}}.)
Topological lensing of the cosmic microwave background (CMB) results in 
multiple images of the same points in different directions.
Pattern formation in the universe's hot and cold
spots reveals the global topology
\cite{{lsdsb},{lbbs}}.
Just as with gravitational lensing,
the location, number and distribution of repeated points will allow
the reconstruction of the geometry.
The circles of Ref. \cite{css} are specific 
collections of topologically lensed
points.  

\begin{figure}
\centerline{
\quad\quad\quad
\psfig{file=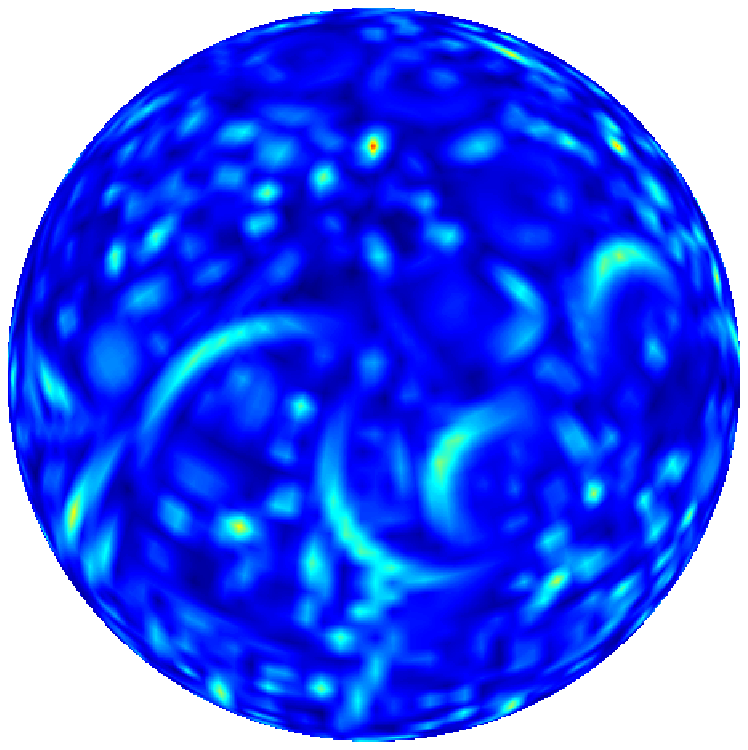,width=2.5in}
} 
\vskip 15truept
\caption{The correlation of every point on the sky with its opposite
in the finite Thurston manifold.}
\label{thurs_ant}
\end{figure}

We demonstrate topological pattern formation 
with the Thurston space, 
popular in homage to the Thurston person \cite{thurs}.
The space corresponds to $m003(-2,3)$ in the {\it SnapPea}
census \cite{snap}.
A CMB map of the sky does not immediately reveal the geometry.
If we scan the sky for correlations between points we can draw out
the hidden pattern.  There are an infinite number of possible
correlated spheres.  The sphere of fig.\ \ref{thurs_ant} is antipody;
the correlation of  
every point on the sky with its opposite point,
	\be
	A(\hat n)=\left <
	{\delta T(\hat n)\over T}{\delta T(-\hat n)\over T}\right >.
	\ee
In an infinite universe, light originating from opposite directions 
would be totally uncorrelated.  The ensemble
average antipodal correlation would produce a monopole with no
structure.
In a finite universe by contrast, light which is received from
opposite directions may in fact have originated from the same location
and simply took different paths around the finite cosmos.
The antipody map would then show structure as it caught the recurrence
of near or identical sources.  Again, the analogy with gravitational
lensing is apparent.

We estimate antipody following the method of
Ref.\ \cite{lsdsb}.
We take the correlation between two points to be
the correlation they would have in an unconnected, infinite space given their
minimum separation.  
The curvature is everywhere negative and the spectrum of fluctuations
are 
taken to be flat and Gaussian, even in the absence of inflation.  
This is justified on a compact, hyperbolic space
since, according to the tenents of quantum chaos, 
the amplitude of quantum fluctuations are drawn
from a Gaussian random ensemble with a flat spectrum consistent with
random matrix theory.
To find the minimum distance
we move the points under comparison back
into 
the fundamental domain using the generators for the compact manifold.  
The result for the Thurston space with $\Omega_o=0.3$ is shown in 
fig.\ \ref{thurs_ant}.
Notice the interesting arcs of correlated points.  Clearly there is
topological lensing at work.  Arcs were also found under antipody for
the Weeks space in Ref.\ \cite{lsdsb}.
If antipody were a symmetry of the space then at least some circles
of correlated points representing the intersection of copies of the
surface of last scatter with itself
would have been located \cite{css}, as were found for
the
Best space \cite{lsdsb}.
Antipody is by definition symmetric under a rotation by $\pi$ and so 
the back of the sphere is identical to the front.

\begin{figure}
\centerline{{\quad\quad\quad
\psfig{file=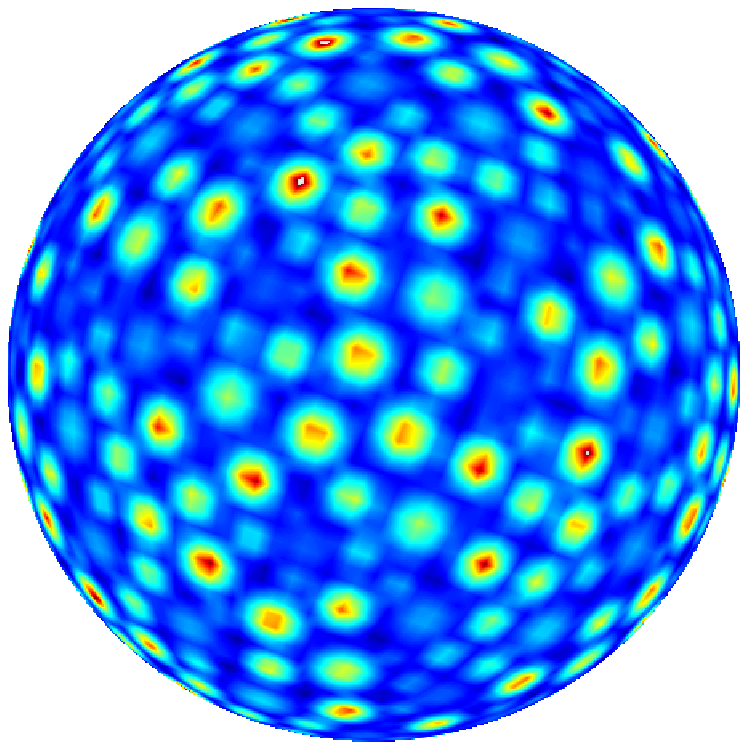,width=2.5in}}} 
\vskip 15truept
\caption{The correlation of one point on the sky with the rest of the
sphere in the Thurston space.  There is a tri-fold symmetry apparent
in the middle of the sphere.
}
\label{origin}
\end{figure}

\begin{figure}
\centerline{{\quad\quad\quad
\psfig{file=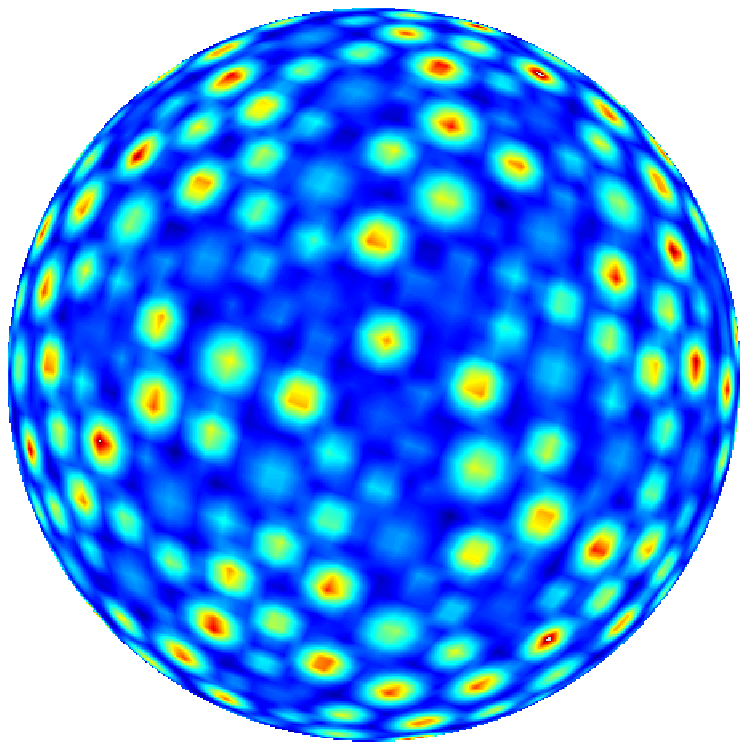,width=2.5in}}} 
\vskip 15truept
\caption{The back of fig.\ \ref{origin}.
The tri-fold symmetry is again apparent
with the three-leaf pattern in the middle of the sphere.
}
\label{origin_back}
\end{figure}

There are an infinite number of correlated spheres which can 
be used to systematically reconstruct the geometry of the fundamental 
domain. Another example is a correlation of one point in the sky with 
the rest of the sphere,
	\be
	C_P(\hat n)=\left
	<{\delta T(\hat n_P)\over T}{\delta T(\hat n)\over T}\right >.
	\ee
This selects out recurrent images of the one 
point.  In an unconnected, infinite space, the sphere would only 
show one spot, namely the
correlation of the point with itself.  In fig.\ \ref{origin} 
we have a kaleidescope of 
images providing detailed information on the underlying space.
There is a trifold symmetry in fig.\ \ref{origin}.  Notice that there
is a band of points moving from the middle upward
vertically which then bends over to the left and that this band repeats twice
making
an overall three-pronged swirl emanating from the middle of the figure.
Since this correlated sphere is not symmetric under $\pi$, we also
show
the back of the sphere in fig.\ \ref{origin_back}.  A different
pattern emerges but still with the tri-fold symmetry.  There is a
three-leaf arrangement of spots in the center of the figure.

We need the improved resolution and signal-to-noise of the future
satellite missions MAP and {\it Planck Surveyor} to observe
topological pattern formation.  High resolution information will be
critical in distinguishing fictitious correlations from real spots.
Beyond the CMB, a finite universe would sculpt the distribution 
of structure on the largest scales.  Even if we never see repeated
images of galaxies or clusters of galaxies, 
the physical distribution of matter
could be shaped by the shape of space.
The topological identifications select discrete modes and the modes
themselves can in turn trace the identifications.  The result is an overall 
web of primordial fluctuations in the gravitational potential 
specific to the finite space.  
A web-like distribution of matter
would then be inherent 
in the initial primordial spectrum \cite{lb}.
This is different from the structureless
distribution of points one would
expect in an infinite cosmos.

We close with the
more fanciful possibility that even time is compact.
If time is compact, every event would repeat precisely as set by the age of
the universe.  
Only a universe which is able to naturally return to its own infancy could be
consistent with a closed time loop.
A big crunch which 
feeds another big bang could allow
our entire history to repeat.   The same galaxies form
and the same stars and planets and people.
Even a proponent of free will can see that at the 
very least we would be limited in the choices we are or are not free to make.
We would live out the same lives, make the same choices, make the same
mistakes.  
Of course, in a quantum creation of the universe,
different galaxies would form in different locations composed of different
stars and new planets.
We would not be here but chances are,
someone would. 
Even if our CMB sky does not look like the Thurston pattern, perhaps
someone's does.

\bigskip
\bigskip

JL thanks the participants and organizers of CTP98.



\begin{thebibliography}{99}

\bibitem{dummy1} Atkatz D.  and Pagels  H., Phys. Rev. D25, 2065 (1982)

\bibitem{dummy2} Birrel N. D.  and Davies  P. C. W., {\sl Quantum Fields in Curved
Spacetime}, Cambridge
Univ. Press, Cambridge, United Kingdom, 1982

\bibitem{dummy3} Bytsenko A. A. and Goncharov Y.,
Class. quantum Grav. 8, 2269, 1991

\bibitem{dummy4} Cornish N. J., Spergel  D. N.  and Starkman G. D. , Phys. Rev. Lett. 77,
215 (1996).

\bibitem{dummy5} Cornish N.,   Spergel D., and Starkman G., Class.Quant.Grav. 15 (1998)
2657-2670; Phys.Rev. D57 (1998) 5982-5996

\bibitem{dummy6} de Oliveira-Costa A.   and Smoot  G., Ap. J. 448, 477 (1995).

\bibitem{dummy7} de Oliveira-Costa A., Smoot  G. and  Starobinsky A., Ap. J. 468, 457, 1996

\bibitem{dummy8} Elizalde E. and Kirsten K., J. Math.  Phys. 35 (3) 1994

\bibitem{dummy9} Ellis G.F. , Q.J.R. Astron. Soc. 16, 245, 1975

\bibitem{dummy10} Goncharov Y.P., Phys. Lett. A 91, 153, 1982

\bibitem{dummy11} Goncharov Y.P.  and Bytsenko  A.A. , Astrophys. 27, 422, 1989


\bibitem{dummy12} Klein O., Zeits. Für Phys., 37, 895, 1926 ; Nature, 118, 516, 1927

\bibitem{dummy13} Lachi{\`e}ze-Rey M., Luminet J.-P., 1995, Phys. Rep. 254,
136 (LaLu)

\bibitem{dummy14} Lehoucq R., Luminet J.-P., Lachi{\`e}ze-Rey M., 1996, A. \& A.
313, 339

\bibitem{dummy15} Lehoucq R., Luminet J.-P.,  Uzan J.-P., A. \& A.  344, 735 (1999)

\bibitem{dummy16} Levin J. J., Barrow J. D., Bunn E. F. and Silk J., Phys. Rev. Lett.
79, 974, 1997

\bibitem{dummy17} Mostepanenko V. M.  and Trunov N. M., Usp. Fiz. Nauk. 156, 385,  1988

\bibitem{dummy18} Nakahara M., {\sl Geometry, Topology and Physics}, Adam Hilger, Bristol 1990

\bibitem{dummy19} Roukema B. F.,  Luminet J.-P., A. \& A. 348 (1999) 8

\bibitem{dummy20} Rovelli C., 1999, preprint /hep-th/9910131

\bibitem{dummy21} Sokolov I.Y., JETP Lett. 57, 617, 1993

\bibitem{dummy22} Spaans M., preprint /arXiv:gr-qc/9901025

\bibitem{dummy23} Starobinsky A.A., JETP Lett. 57, 622, 1993

\bibitem{dummy24} Stevens D., Scott  D.  and Silk J., Phys. Rev. Lett. 71, 20, 1993

\bibitem{dummy25} Souriau J. - M., Nuovo cimento, XXX, 2,  1963

\bibitem{dummy26} Thiry Y., Journal Math. Pures et Apppl., 9, 1 (1947)

\bibitem{dummy27} Thurston W. P., {\sl The
Geometry and Topology of 3-Manifolds}, Princeton University Press, Princeton,
1978

\bibitem{dummy28} Thurston W. P., Bull. Am. Math. Soc. 6, 357, 1982

\bibitem{dummy29} Zel'dovich Ya. B.  and Starobinsky  A. A., Sov. Astron. Lett. 10, 135 (1984)
\end{thebibliography}

\begin{thebibliography}{9} \parskip=-9pt
\bibitem{FeC}  H. V. Fagundes and S. S. e Costa, preprint arXiv:gr-qc/9801066, to
appear in Gen. Relat. Gravit. 31 (1999).

\bibitem{GWG}  G. W. Gibbons, Class. Quantum Grav. 15, 2605 (1998)

\bibitem{LaLu}  M. Lachi\`{e}ze-Rey and J.-P. Luminet, Physics Rep. 254, 135
(1995)

\bibitem{Scott}  P. Scott, Bull. London Math. Soc. 15, 401 (1983)

\bibitem{DeLor}  V. A. De Lorenci, J. Martin, N. Pinto-Neto, and I. D.
Soares, Phys. Rev. D56, 3329 (1997)
\end{thebibliography}

\begin{thebibliography}{99}

\bibitem[Bond, Pogosyan \& Souradeep(1998)]{BPS98} \joref{Bond J.~R., Pogosyan D., Souradeep T.}{\cqg}{15}{2573}{1998}\ (arXiv:astro-ph/9804041)





\bibitem[Cornish, Spergel \& Starkman(1996)]{Corn96} Cornish N.~J., Spergel D.~N., Starkman G.~D., 1997, arXiv:gr-qc/9602039
\bibitem[Cornish, Spergel \& Starkman(1998b)]{Corn98b} \joref{Cornish N.~J., Spergel D.~N., Starkman G.~D.}{\cqg}{15}{2657}{1998b}\ (arXiv:astro-ph/9801212) 


%
\bibitem[Lachi\`eze-Rey \& Luminet(1995)]{LaLu95} \joref{Lachi\`eze-Rey M.,  Luminet J.-P.}{PhysRep}{254}{136}{1995}
\bibitem[Lehoucq et {al.}(1996)]{LLL96} \joref{Lehoucq R., Luminet J.-P., Lachi\`eze-Rey M.}{\aanda}{313}{339}{1996}
%
%
%
\bibitem[Luminet(1998)]{Lum98} \epref{Luminet J.-P.}{arXiv:gr-qc/9804006}{1998}
\bibitem[Luminet \& Roukema(1999)]{LR99} {Luminet J.-P., Roukema B.~F.}, 1999, in {\em Theoretical and Observational Cosmology, NATO Advanced Study Institute, Carg\`ese 1998,} ed. Lachi\`eze-Rey, M., Netherlands:Kluwer, p117 \ (arXiv:astro-ph/9901364)

%
%
\bibitem[Perlmutter et {al.}(1999)]{SCP9812} \joref{Perlmutter S. et al.}{\apj}{517}{565}{1999} ~(arXiv:astro-ph/9812133)

%
%
%
\bibitem[Roukema(1996)]{Rouk96} \joref{Roukema B.~F.}{\mnras}{283}{1147}{1996}
\bibitem[Roukema(2000a)]{Rouk99} Roukema B.~F., 2000a, \mnras, 312, 712 \ (arXiv:astro-ph/9910272)
\bibitem[Roukema(2000c)]{Rouk00c} Roukema B.~F., 2000b, \cqg, 17, 3951 \ (arXiv:astro-ph/0007140)
\bibitem[Roukema \& Bajtlik(1999)]{RBa99} Roukema B.~F., Bajtlik, S., 1999, \mnras, 308, 309 \ (arXiv:astro-ph/9901299)
\bibitem[Roukema \& Edge(1997)]{RE97} \joref{Roukema B.~F., Edge A.~C.}{\mnras}{292}{105}{1997}
\bibitem[Starkman(1998)]{Stark98} \joref{Starkman G.~D.}{\cqg}{15}{2529}{1998}

\bibitem[Uzan et al.(1999)Uzan, Lehoucq \& Luminet(1999)]{ULL99a} Uzan J.-Ph., Lehoucq R., Luminet J.-P., 1999, \aanda, 351, 766  \ (arXiv:astro-ph/9903155)
%

\end{thebibliography}

\begin{thebibliography}{99}

\bibitem{lalu}M. Lachi\`eze-Rey and J.-P. Luminet, Phys. Rep. {\bf 254}, 
136 (1995), (arXiv:gr-qc/9605010). 

\bibitem{lumi}J.-P. Luminet, arXiv:astro-ph/9804006. 

\bibitem{wei}S. Weinberg, {\it Gravitation and Cosmology} 
(John Wiley, New York, 1972). 

\bibitem{tu98}M. S. Turner, {\it Cosmology Update 1998}, 
arXiv:astro-ph/9901168.

\bibitem{rolu}B. F. Roukema and J.-P. Luminet, arXiv:astro-ph/9903453.

\bibitem{cornetal} N. J. Cornish, D. Spergel and G. Starkman, 
Phys. Rev. D57, 5982 (1998), (arXiv:astro-ph/9708225).

\bibitem{luro}J.-P. Luminet and B. F. Roukema, Proceedings 
of Cosmology School held at Cargese, Corsica, 
August 1998, arXiv:astro-ph/9901364. 

\bibitem{dela}M. Demia\'nski and M. Lapucha, Mon. Not. R. astr. Soc. 
{\bf 224}, 527 (1987). 

\bibitem{fawi}H. V. Fagundes and U. F. Wichoski, Astrophys. 
J. {\bf 322}, L5 (1987).

\bibitem{ro}B. F. Roukema, Mon. Not. R. astr. Soc. {\bf 283}, 
1147 (1996), (arXiv:astro-ph/9603052).

\bibitem{robla}B. F. Roukema and V. Blanloeil, Class. Quant. 
Grav. {\bf 15} 2645 (1998), (arXiv:astro-ph/9802083).

\bibitem{supernova}S. Perlmutter {\it et al.}, arXiv:astro-ph/9812133.

\end{thebibliography}

\begin{thebibliography}{99}

\bb{Sokolov}
I. Yu., Sokolov, JETP Lett, {\bf 57} 10, 621 (1993)

\bb{Stevens}
D. Stevens, D. Scott and J. Silk, Phys. Rev. Lett.
{\bf 71}, 20 (1993) 

\bb{Oliveira}
A. de Oliveira-Costa, G.F. Smoot and A.A. Starobinsky, Astrophys. J. 
{\bf 468}, 457 (1996)  

\bb{Levin}
J.L. Levin, E. Scannapieco and J Silk, Phys. Rev. D 
{\bf 58}, 103516 (1998)

\bb{Roukema}
B.F. Roukema, Class. Quantum. Grav. {\bf 17}, 3951 (2000)

\bb{Cornish1}
N. Cornish, D. Spergel and G. Starkman, Phys. Rev. D {\bf 57},
5982 (1998)


\bb{Cornish2} 
N.J. Cornish and D.N. Spergel, Phys. Rev. D 
(in press) (astro-ph/9906401)


\bb{Inoue1}
K.T. Inoue, K. Tomita and N. Sugiyama, MNRAS {\bf 314}, No.4,1, L21
(2000a)


\bb{Aurich}
R. Aurich, Astrophys. J. {\bf 524}, 497 (1999)


\bb{Inoue2}
K.T. Inoue, Phys. Rev. D {\bf 62}, 103001 
 (2000b)

\bb{Bond1}
J.R. Bond, D. Pogosyan \& T. Souradeep, 
Class.Quant.Grav. 15 2671 (1998) 

\bb{Bond2}
J.R. Bond, D. Pogosyan \& T. Souradeep, 
Phys. Rev. D {\bf 62} 043006 (2000)

\bb{Inoue3}
K.T. Inoue, in preparation (2000c)

\end{thebibliography}

\begin{thebibliography}{99}

\bibitem{jin} J.J.Levin, {\it Phys. Lett.} {\bf B 343 } (1995) 69.

\bibitem{lum} M. Lachieze-Rey and J.P. Luminet, {\em Phys. Rep.} {\bf 254}
(1995) 135.


\bibitem{volume} The entire volume
{\it Class. Quantum Grav.} {\bf 15}
(1998) 2671; for other CMB approaches see also
J.R. Bond, D. Pogosyan and T. Souradeep,
arXiv:astro-ph/9702212 (1997).

\bibitem{lsdsb}
J. Levin, E. Scannapieco, G. de Gasperis, J. Silk and
J.D. Barrow,
{\em Phys. Rev. } { D} {\bf 58} (1998) article 123006.


\bibitem{lbbs}  J. Levin, J.D. Barrow, E.F. Bunn and J. Silk, {\em Phys.
Rev. Lett.} {\bf 79} (1997) 974.


\bibitem{css} 
N.J. Cornish, D. Spergel and G. Starkman, 
{\it Phys. Rev. Lett.} {\bf 77}, 215 (1996);  
{\it ibid.}
{\em Phys. Rev. } {\bf D57} (1998) 5982.


\bibitem{thurs}
W.P. Thurston, Bull. Am. Math. Soc. {\bf 6} (1982) 357;
W.P. Thurston and J.R. Weeks, Sci. Am. July (1984) 94;
W.P. Thurston, ``Three-dimensional geometry and topology''
(Ed: Silvio Levy, Princeton University Press, Princeton, N.J. 1997).

\bibitem{snap}  J. Weeks, computer software {\it SnapPea} available at
http://www.geom.umn.edu:80/software.

\bibitem{lb} J. Levin and J.D. Barrow, 
in preparation; {\it ibid} conference proceedings for ``The Chaotic
Universe'' ICRA, Rome 1999.


\end{thebibliography}
\end{document}